\documentstyle[proc,epsf]{rspublic}

\newcommand{\la}{\langle} 
\newcommand{\ra}{\rangle} 
\newtheorem{guess}{Proposition}  
\newtheorem{lem}{Lemma} 
 
\newtheorem{theo}{Theorem} 

\def\thedemobiblio#1{\smallskip\par
 \list{}{\labelwidth 0pt \leftmargin 1em \itemindent -1em \itemsep 1pt}
 \small \parindent 0pt
 \parskip 1.5pt plus .1pt\relax
 \def\newblock{\hskip .11em plus .33em minus .07em}
 \sloppy\clubpenalty4000\widowpenalty4000
 \sfcode`\.=1000\relax}

\begin{document}

\title[Statistical Geometry in Quantum Mechanics]
{Statistical Geometry in Quantum Mechanics} 

\author[D.C. Brody, L.P. Hughston]{
Dorje C. Brody$^{*}$ 
and 
Lane P. Hughston$^{\dagger}$  
} 
\affiliation{$*$Blackett Laboratory, Imperial College, 
South Kensington, London SW7 2BZ \\ 
and DAMTP, Silver Street, Cambridge CB3 9EW, U.K. \\ 
$\dagger$ Merrill Lynch International, 
25 Ropemaker Street, London EC2Y 9LY U.K. \\ 
and King's College London, The Strand, London 
WC2R 2LS, U.K.}


\maketitle 
\input{psfig.sty}
 
\begin{abstract} 
A statistical model ${\cal M}$ is a family of probability 
distributions, characterised by a set of continuous parameters known 
as the parameter space. This possesses natural geometrical properties 
induced by the embedding of the family of probability distributions 
into the space of all square-integrable functions. More precisely, by 
consideration of the square-root density function we can regard 
${\cal M}$ as a submanifold of the unit sphere ${\cal S}$ in a real 
Hilbert space ${\cal H}$. Therefore, ${\cal H}$ embodies the `state 
space' of the probability distributions, and the geometry of the given 
statistical model can be described in terms of the embedding of 
${\cal M}$ in ${\cal S}$. The geometry in question is characterised 
by a natural Riemannian metric (the Fisher-Rao metric), thus allowing 
us to formulate the principles of classical statistical inference in a 
natural geometric setting. In particular, we focus attention on the 
variance lower bounds for statistical estimation, and establish 
generalisations of the classical Cram\'er-Rao and Bhattacharyya 
inequalities, described in terms of the geometry of the underlying 
real Hilbert space. As a comprehensive illustration of the utility of 
the geometric framework, the statistical model ${\cal M}$ is then 
specialised to the case of a submanifold of the state space of a 
quantum mechanical system. This is pursued by introducing a 
compatible complex structure on the underlying real Hilbert space, 
which allows the operations of ordinary quantum mechanics to be 
reinterpreted in the language of real Hilbert space geometry. The 
application of generalised variance bounds in the case of quantum 
statistical estimation leads to a set of higher order corrections to 
the Heisenberg uncertainty relations for canonically conjugate 
observables. \par 
\end{abstract}

\section{Introduction} 

The purpose of this paper is twofold: first, to develop a concise 
geometric formulation of statistical estimation theory; and second, 
the application of this formalism to quantum statistical inference. 
Our intention is to establish the basic concepts of statistical 
estimation within the framework of Hilbert space geometry. This line 
of enquiry, although suggested by Bhattacharyya (1942), Rao (1945), 
and Dawid (1975,1977), has not hitherto been pursued in the spirit of 
the fully geometric program that we undertake here. In 1945 Rao 
introduced a Riemannian metric, in local coordinates given by the 
components of the Fisher information matrix, on the parameter space 
of a family of probability distributions. He also introduced the 
corresponding Levi-Civita connection associated with the Fisher 
information matrix, and proposed the geodesic distance induced by 
the metric as a measure of dissimilarity between probability 
distributions. Thirty years after Rao's initial work, Efron (1975) 
carried the argument a step forward when he introduced, in effect, 
a new affine connection on the parameter space manifold, and thus 
shed light on the role of the embedding curvature of the statistical 
model in the relevant space of probability distributions. The work 
of Efron has been followed up and extended by a number of authors 
(see, e.g., Amari 1982, 1985, Barndorff-Nielsen, Cox, and Reid 1986, 
and Kass 1989), particularly in the direction of asymptotic 
inference. However, the applicability of modern differential 
geometric methods to statistics remains in many respects a 
surprising development, about which there is still much to be 
learned.  \par 

In a remark on Efron's construction, Dawid (1975) asked whether there 
might be a fundamental role played by the Levi-Civita connection in 
statistical analysis. The aim of this paper in part is to answer 
this important question, by studying statistical inference from a 
Hilbert space perspective. In particular, we shall study the 
geometric properties of a statistical model ${\cal M}$ induced when 
we embed ${\cal M}$ via the square-root map in the unit sphere 
${\cal S}$ in a real Hilbert space ${\cal H}$. This leads in a 
natural way to the Levi-Civita connection on ${\cal M}$. \par 

It was also pointed out by Dawid (1977), in the case of an embedding 
given by the square-root of the likelihood function, that the 
Hilbert space norm induces a spherical geometry (see also Burbea 
1986). If the density function is parameterised by a set of 
parameters $\theta$, then for each value of $\theta$ we have a 
corresponding point on the unit sphere $\cal{S}$ in the Hilbert 
space $\cal{H}$. By choosing a basis in $\cal{H}$, we can 
associate a unit vector $\xi^{a}(\theta)$ with this point, and work 
with the abstract vector $\xi^{a}(\theta)$ instead of 
$\sqrt{p_{\theta}(x)}$. The index `$a$' is abstract in the sense 
that we do not necessarily regard it as `taking values'; instead, 
it serves as a kind of `place-keeper' for various tensorial 
operations. We show how the abstract index approach can be used as 
a powerful tool in statistical investigations. \par 

Our program includes the exploitation of this methodology to study 
geometrical and statistical aspects of quantum mechanics. The 
specialisation to quantum theory requires an extra ingredient, 
namely, a {\it complex structure}. Thus, if we take our real Hilbert 
space and impose on it a complex structure, compatible with the real 
Hilbert space metric, the resulting geometry is sufficiently rich to 
allow us to introduce all of the standard operations of quantum 
theory. \par 

While the conventional approach to quantum statistical 
estimation has essentially been merely `by analogy' with classical 
estimation, our approach differs in the sense that we view quantum 
estimation theory as arising in essence as a natural extension 
of the classical theory, when the theory is `enriched' with the 
addition of a complex structure, and the system of random variables 
is expanded to include incompatible observables. \par 

By way of contrast 
we note that most of the current literature of quantum statistical 
estimation (see, e.g., Accardi and Watson 1994, Braunstein and Caves 
1994, Brody and Meister 1996a,b, Helstrom 1976, Holevo 1982, 
Ingarden 1981, Jones 1994, Malley and Hornstein 1993, Nagaoka 1994, 
and references cited therein) takes the space of density matrices as 
the relevant state space in terms of which estimation problems are 
formulated, the view there being that the `space of density matrices' 
is the quantum mechanical analogue of the `space of density 
functions' when we consider the quantum estimation problem. \par 

In our approach, however, we find it useful to emphasise the role 
of the space of {\it pure} quantum states. In fact, the space of 
density matrices has a very complicated geometric structure, owing 
essentially to the various levels of `degeneracy' a density matrix 
can possess, and the relation of these levels to one another. It can 
be argued that to tackle the quantum estimation problem head-on from 
a density matrix approach is not necessarily advantageous. In any 
case, the consideration of pure states allows us to single out most 
sharply the relation between classical statistical theory and quantum 
statistical theory, and in such a way that the geometry takes on a 
satisfactory character. The extension of our approach to general 
states will be taken up elsewhere. \par 

The plan of the paper is as follows. In \S 2, the geometry of the 
parameter space induced by the Hilbert space norm is introduced by 
means of an index notation. This notation is employed here 
partially for the purpose of simplifying complicated calculations, 
and its usefulness in this respect will become evident. The index 
notation also greatly facilitates the geometrical interpretation of 
the operations being represented here. Attention is drawn to 
formula (\ref{eq:f-inf}) for the Riemannian metric on ${\cal M}$, 
and the argument given in Proposition \ref{prop:12} that indicates 
the special status of the Levi-Civita connection. Our idea is to 
reformulate a number of the standard concepts of statistics in the 
language of Hilbert space geometry. In particular, in \S 3 and \S 
4 we develop the theory of the maximum likelihood estimator (MLE) 
and the Cram\'er-Rao (CR) variance lower bound, for which novel 
geometrical interpretations are provided. See, for example, 
Proposition \ref{prop:mle} and Theorem 1. Also, note Proposition 
\ref{prop:21} where a striking link is made between an essentially 
statistical quantity and an essentially geometric quantity. In \S 
4 we consider in some detail properties of the canonical family of 
exponential distributions, which can be described concisely in 
terms of the Hilbert space geometry. This material also has 
interesting applications to statistical mechanics and thermal 
physics, which we discuss elsewhere. \par 

In \S 5, a set of higher-order 
corrections to the CR lower bound is obtained, leading to what 
might appropriately be called generalised Bhattacharyya bounds, 
given in Proposition \ref{prop:Bhat}. However, unlike the original 
Bhattacharyya bound, the new variance bounds generally depend upon 
features of the estimator. Nevertheless, in certain cases of 
interest the result is independent of the specific choice of 
estimator. This will be illustrated with examples from problems in 
quantum estimation. A brief account of multi-parameter situation 
is given in \S 6. \par 

After some comments on the transition from classical to quantum 
theory in \S 7, a general geometric formulation of ordinary quantum 
mechanics is developed in \S8 and \S9 in terms of a real Hilbert 
space setting. In \S 10 and \S 11 we apply our geometric estimation 
theory to the quantum mechanical state space. We are interested, in 
particular, in the variance bounds associated with pairs of 
canonically conjugate observables. Here we study in detail the 
example of time estimation in quantum theory. This is pursued by 
means of a nonorthogonal resolution of unity, known as a probability 
operator-valued measure (POM), which allows 
us to construct a well defined maximally symmetric time 
`observable' within the framework of ordinary quantum mechanics. 
Finally in \S 12, we apply the generalised variance lower bounds 
to obtain a remarkable set of higher order corrections to the 
Heisenberg relations. \par

\section{Index Notation and Fisherian Geometry} 

Consider a real Hilbert space $\cal{H}$, equipped with a symmetric 
inner product which we denote $g_{ab}$. As noted above, we adopt an 
index notation for Hilbert space operations. Let us write $\xi^{a}$ 
for a typical vector in $\cal{H}$. If $\cal{H}$ is finite, the 
index can be thought of as ranging over a set of integers, while 
in the infinite dimensional case, the index is `abstract'. See 
Geroch (1971a,b), Penrose and Rindler (1984, 1986), or Wald (1994) 
for further details of this notation. 
Our intention here is not to present a rigourous account of the 
matter, which would be beyond the scope of the present work, but 
rather to illustrate the utility of the index calculus by way of a 
number of examples. In particular, in the infinite dimensional 
case there are technical conditions concerning the domains of 
operators that require care---these will not concern us here in 
the first instance, though in our treatment of quantum estimation 
more attention will be paid in this respect. \par 

Suppose we consider the space of all probability density functions 
$p(x)$ on the sample space ${\bf R}^{n}$. By taking a square root 
we can map each density function to a point on the unit sphere 
$\cal{S}$ in ${\cal H}=L^{2}({\bf R}^{n})$, given by 
$g_{ab}\xi^{a}\xi^{b} = 1$. A random variable in $\cal{H}$ is then 
represented by a symmetric bilinear form, e.g., $X_{ab}$, with 
expectation $X_{ab}\xi^{a}\xi^{b}$ in the state $\xi^{a}$, that is, 
\begin{equation} 
E_{\xi}[X]\ =\ X_{ab}\xi^{a}\xi^{b}\ . 
\end{equation} 
In terms of the conventional statistical notation, we can associate 
$\xi^{a}$ with $p(x)^{1/2}$, $X_{ab}$ with $x\delta(x-y)$, and hence 
$X_{ab}\xi^{a}\xi^{b}$ with the integral 
\begin{equation} 
\int_{x}\int_{y}x\delta(x-y) p(x)^{1/2}p(y)^{1/2}dxdy\ , 
\end{equation} 
which reduces to the expectation. This line of reasoning can be 
extended to more general expressions. Thus, for example, 
$X_{ab}X^{b}_{c}\xi^{a}\xi^{c}$ is the expectation of the square of 
the random variable $X_{ab}$, and the variance of $X_{ab}$ in 
the state $\xi^{a}$ is 
\begin{equation} 
{\rm Var}_{\xi}[X]\ =\ \tilde{X}_{ac}\tilde{X}^{c}_{b}
\xi^{a}\xi^{b}\ , 
\end{equation} 
where $\tilde{X}_{ab} = X_{ab}-g_{ab}(X_{cd}\xi^{c}\xi^{d})$ 
represents the deviation $\Delta X$ of the random variable from its 
mean. Likewise for the covariance of the random variables $X_{ab}$ 
and $Y_{ab}$ in the state $\xi^{a}$ we can write 
\begin{equation} 
{\rm Cov}_{\xi}[X,Y]\ =\ {\tilde X}_{ac}{\tilde Y}^{c}_{b} 
\xi^{a}\xi^{b}\ . 
\end{equation} 
Note that if $\xi^{a}$ is not normalised, then the formulae 
above can be generalised with the inclusion of suitable normalisation 
factors. \par 

We consider now the unit sphere $\cal{S}$ in $\cal{H}$, and within 
this sphere a submanifold $\cal{M}$ given parametrically by 
$\xi^{a}(\theta)$, where $\theta^{i}\ (i=1, \cdots,r)$ are local 
parameters. We write $\partial_{i}$ for $\partial/\partial\theta^{i}$. 

\begin{guess} 
${\rm (}$Fisher-Rao metric${\rm )}$. 
In local coordinates, the Riemannian metric ${\cal G}_{ij}$ 
on $\cal{M}$, induced by $g_{ab}$, given by 
\begin{equation} 
{\cal G}_{ij}\ =\ 4 g_{ab}\partial_{i}\xi^{a} 
\partial_{j}\xi^{b} \ , \label{eq:f-inf} 
\end{equation} 
is the Fisher information matrix. 
\label{prop:11} 
\end{guess} 

The proof is as follows. We note that the squared distance 
between the endpoints of two vectors $\xi^{a}$ and $\eta^{a}$ in 
$\cal{H}$ is $D^{2} = g_{ab}(\xi^{a}-\eta^{a})(\xi^{b}-\eta^{b})$. 
If both endpoints lie on $\cal{M}$, and $\eta^{a}$ is obtained by 
infinitesimally displacing $\xi^{a}$ in $\cal{M}$, i.e., 
$\eta^{a}=\xi^{a}+\partial_{i}\xi^{a}d\theta^{i}$, then the 
separation $ds$ between the two endpoints on $\cal{M}$ is 
\begin{equation}  
ds^{2}\ =\ \frac{1}{4} {\cal G}_{ij}d\theta^{i}d\theta^{j}\ , 
\end{equation} 
where ${\cal G}_{ij}$ is given as in (\ref{eq:f-inf}). The factor of 
$\frac{1}{4}$ arises from the conventional definition of the Fisher 
information matrix in terms of the log-likelihood function 
$l(x|\theta)=\ln p(x|\theta)$, given by 
\begin{equation} 
{\cal G}_{ij}\ =\ \int_{x} 
p(x|\theta)\partial_{i}l(x|\theta)\partial_{j}l(x|\theta)dx\ . 
\end{equation} 
By differentiating $g_{_ab}\xi^{a}\xi^{b}=1$ twice, we obtain an 
alternative expression for ${\cal G}_{ij}$, that is, 
${\cal G}_{ij} = - 4 \xi_{a}\partial_{i}\partial_{j}\xi^{a}$. 
This formula turns out to be useful in statistical mechanics (see, 
e.g., Brody and Rivier 1995, Streater 1996), where the geometry of 
the relevant coupling constant space can be investigated. The 
induced geometry of $\cal{M}$ can be studied in terms of the metric 
${\cal G}_{ij}$ and our subsequent analysis will be pursued on this 
basis. To start, we note the following result: 

\begin{lem} 
The Christoffel symbols for the metric connection arising from 
$\cal{G}$ are given by $\Gamma^{i}_{jk} = 
4 {\cal G}^{il}\partial_{l}\xi^{a}\partial_{j}\partial_{k}\xi_{a}$. 
\label{lem:11} 
\end{lem} 
This can be obtained by insertion of (\ref{eq:f-inf}) into 
the familiar formula 
\begin{equation}  
\Gamma^{i}_{jk}\ =\ \frac{1}{2} {\cal G}^{il} 
\left( \partial_{j}{\cal G}_{kl} + \partial_{k}{\cal G}_{jl} 
- \partial_{l}{\cal G}_{jk} \right) 
\end{equation} 
for the Levi-Civita (metric) connection. Now, let $\nabla_{i}$ denote 
the standard Levi-Civita covariant derivative operator associated 
with ${\cal G}_{ij}$, for which $\nabla_{i}{\cal G}_{jk}=0$ and 
$\nabla_{i}$ is torsion free. Then a straightforward calculation 
shows that 
\begin{equation} 
{\cal G}_{ij}\ =\ -4\xi_{a}\nabla_{i}\nabla_{j}\xi^{a}\ . 
\label{eq:fish2} 
\end{equation} 
A question that naturally arises is, 
are there any other `natural' connections associated with the given 
Hilbert space structure? This requires one to construct a tensor of 
the form $Q_{ijk}$ purely from the metric and covariant derivatives 
of the state vector $\xi^{a}$. The answer to this question is of 
relevance, since we would like to know whether it is possible to 
construct a set of affine connections (e.g., Amari's 
$\alpha$-connection) purely in terms of the given basic Hilbert 
space geometry, or whether extra structure is required. Clearly, the 
only possibilities are $\nabla_{i}\xi^{a}\nabla_{j}\nabla_{k}\xi_{a}$ 
and $\xi^{a}\nabla_{i}\nabla_{j}\nabla_{k}\xi_{a}$. However, some 
straightforward algebra leads us to the following result. \par 

\begin{guess} 
The expressions $\nabla_{i}\xi^{a}
\nabla_{j}\nabla_{k}\xi_{a}$ $and$ $\xi^{a}\nabla_{i}
\nabla_{j}\nabla_{k}\xi_{a}$ vanish. Thus, no natural 
three-index tensors can be constructed in Hilbert space, 
and the Levi-Civita connection is distinguished amongst 
possible $\alpha$-connections.  
\label{prop:12} 
\end{guess} 

The proof is as follows. 
First, note that $\nabla_{k}{\cal G}_{ij}=0$ implies 
$\nabla_{k}(\nabla_{i}\xi^{a}\nabla_{j}\xi_{a}) = 0$, and 
hence $\nabla_{k}\nabla_{i}\xi^{a}\nabla_{j}\xi_{a} 
+ \nabla_{k}\nabla_{j}\xi^{a}\nabla_{i}\xi_{a} = 0$. 
On the other hand, it follows from (\ref{eq:fish2}), by 
differentiation, that 
$\nabla_{k}\xi_{a}\nabla_{i}\nabla_{j}\xi^{a} = -\xi_{a} 
\nabla_{k}\nabla_{i}\nabla_{j}\xi^{a}$. 
Therefore, we deduce that 
$\nabla_{k}\nabla_{i}\xi^{a}\nabla_{j}\xi_{a} = 
\xi_{a}\nabla_{i}\nabla_{j}\nabla_{k}\xi^{a}$, 
and that 
$\xi_{a}\nabla_{(i}\nabla_{j)}\nabla_{k}\xi^{a} = 0$. 
Since $\xi_{a}\nabla_{i}\nabla_{j}\nabla_{k}\xi^{a}$ is 
antisymmetric over the indices $i,j$, it follows that 
\begin{equation}  
\xi_{a}\nabla_{i}\nabla_{j}\nabla_{k}\xi^{a}\ =\ 
\frac{1}{2}\xi_{a}R_{ijk}^{\ \ \  l}\nabla_{l}\xi^{a} \label{eq:38} 
\end{equation}  
where $R_{ijk}^{\ \ \  l}$ is the Riemann tensor, defined by 
$(\nabla_{i}\nabla_{j}-\nabla_{j}\nabla_{i})V_{k} = 
R_{ijk}^{\ \ \  l} V_{l}$  
for any smooth vector field $V_{k}$. However, 
$\xi_{a}\nabla_{l}\xi^{a}$ vanishes in (\ref{eq:38}), since 
$\xi^{a}\xi_{a}=1$, and that establishes the desired result.\par 

Therefore, to introduce other affine connections on ${\cal M}$, such 
as Amari's $\alpha$-connection, additional structure on the given 
Hilbert space is required. Although these `artificial' connections 
are useful in certain statistical inference problems, such as higher 
order asymptotics, we conclude that from a Hilbert space point of 
view the Levi-Civita connection is the only `natural' connection 
associated with the space of probability measures. \par 

Note, incidentally, that in the case of a one-parameter family of 
distributions, the Fisher information is given by 
${\cal G} = 4g_{ab}\dot{\xi}^{a}\dot{\xi}^{b}$, 
where the dot denotes differentiation with respect to $\theta$. 
Thus, the Fisher information is related in a simple way to the 
`velocity' along the given curve in 
Hilbert space. This is a result that, as we see later (cf. 
Proposition \ref{lem:22}), has profound links with an analogous 
construction in quantum mechanics (Anandan and Aharonov 1990). \par 

\section{Maximum Likelihood Estimation} 

Suppose we are given a random variable $X_{ab}$ which takes real 
values, and told that the result of a sampling of $X_{ab}$ is the 
number $x$. We are interested in a situation where we have a 
one-parameter family of normalised states $\xi^{a}(\theta)$ 
characterising the distribution of $x$. The parameter $\theta$ 
determines the unknown state of nature, and we wish to estimate 
$\theta$ by use of maximum likelihood methods; that is, we wish to 
associate with a given value of $x$ an appropriate value of $\theta$ 
that maximises the likelihood function. In this section, we present 
a geometrical characterisation of the maximum likelihood estimator 
(MLE), which has an elegant Hilbert space interpretation once we 
single out a `preferred' random variable $X_{ab}$. \par 

\begin{guess} 
Given the random variable $X_{ab}$, the normalised state vector 
$\xi^{a}(\theta)$, and the measurement outcome $x$, the 
parameterised likelihood function $p(x|\theta)$ is given by 
\begin{equation} 
p(x|\theta)\ =\ \frac{1}{\sqrt{2\pi}} 
\int_{-\infty}^{\infty} \xi^{a}\xi_{b} 
\exp\left[ i\lambda(X_{a}^{b} - 
x\delta_{a}^{b})\right] d\lambda\ . \label{eq:likf} 
\end{equation} 
\label{prop:13} 
\end{guess} 

This can be seen as follows. We define the projection 
operator $\Delta^{a}_{b}$ associated with the random variable 
$X_{ab}$ and the measurement outcome $x$ by  
\begin{equation}  
\Delta^{a}_{b}(X,x)\ =\ \frac{1}{\sqrt{2\pi}} 
\int_{-\infty}^{\infty} \exp\left[ i\lambda(X^{a}_{b} 
- x\delta^{a}_{b})\right]d\lambda\ . 
\end{equation} 
We note that $\Delta^{a}_{c}(X,x)\Delta^{c}_{b}(X,y) = 
\Delta^{a}_{b}(X,x)\delta(x-y)$, and that $X_{ab}$ can be recovered 
from $\Delta_{ab}(X,x)$ via the spectral resolution 
\begin{equation} 
X_{ab}\ =\ \int_{-\infty}^{\infty}x\Delta_{ab}(X,x)dx\ . 
\end{equation} 
Then, the likelihood function $p(x|\theta)$ is the expectation of 
$\Delta_{ab}$ in the state $\xi^{a}$, i.e., 
\begin{equation}  
p(x|\theta)\ =\ \Delta_{ab}\xi^{a}\xi^{b}\ . \label{eq:trace} 
\end{equation} 
Alternatively, $p(x|\theta)$ can be obtained by taking the 
Fourier transform of the characteristic function 
\begin{equation}  
\phi_{\theta}(\lambda)\ =\ \xi^{a}\xi_{b} 
\exp\left[ i\lambda X^{b}_{a}\right] \ , 
\end{equation}  
which leads back to (\ref{eq:likf}). The maximum likelihood 
estimate $\bar{\theta}(x)$ for $\theta$, assuming it exists 
and is unique, is obtained by solving 
\begin{equation} 
\Delta_{ab}(X,x) \xi^{a}
\dot{\xi}^{b}\ _{|\theta=\bar{\theta}} \ =\ 0\ . \label{eq:mle} 
\end{equation} 
Geometrically, this means that, along the curve $\xi^{a}(\theta)$ on 
the sphere ${\cal S}$, $\bar{\theta}(x)$ maximises the quadratic 
form $\Delta_{ab}\xi^{a}\xi^{b}$. Conversely, if $\bar{\theta}(x)$ 
is the MLE for the parameter $\theta$, then the corresponding random 
variable in $\cal{H}$ is 
\begin{equation} 
\Theta_{ab}(X)\ =\ 
\int_{-\infty}^{\infty} \bar{\theta}(x) 
\Delta_{ab}(X,x)dx\ . 
\end{equation} 
If we let $\Delta_{x}(\xi^{a})$ denote the quadratic form 
$\Delta_{ab}\xi^{a}\xi^{b}$ on ${\cal H}$, then equation 
(\ref{eq:mle}) for the MLE can be rewritten as 
$\dot{\xi}^{a}\nabla_{a}\Delta_{x} = 0$, where the `gradient' 
operator $\nabla_{a}$ is defined by $\nabla_{a}=\partial/\partial 
\xi^{a}$, so $\nabla_{a}(\Delta_{bc}\xi^{b}\xi^{c}) = 
2\Delta_{ab}\xi^{b}$. Thus, for each 
given value of $x$ we can foliate $\cal{S}$ with hypersurfaces 
of constant $\Delta_{x}$. This leads us to the following 
characterisation of the MLE. 

\begin{guess} 
The maximum likelihood estimate $\bar{\theta}(x)$ is the value of 
$\theta$, for each given value of $x$, such that the tangent of the 
curve $\xi^{a}(\theta)$ is orthogonal to the normal vector of the 
constant $\Delta_{x}$ surface passing through the point 
$\xi^{a}(\theta)$. 
\label{prop:mle} 
\end{guess} 

Thus, we see that maximum likelihood estimation has a 
characterisation in terms of Hilbert space geometry that can be 
achieved by introducing extra structure on ${\cal H}$, namely, 
by `singling out' a particular observable. This is natural in the 
context of some classical statistical investigations, though for 
quantum statistical inference we may wish to avoid the introduction 
of `preferred' observables. \par 

\section{Cram\'er-Rao Lower Bound and Exponential Families} 

In the case of a general estimation problem, a lower bound can 
be established for the variance with which the estimate deviates 
from the true value of the relevant parameter. Our intention in 
this section is to present a geometric characterisation of this 
bound. In doing so we also make some observations about the geometry 
of exponential families of distributions, of relevance to statistical 
physics. Consider a curve $\xi^{a}(\theta)$ in $\cal{S}$. We say that 
a random variable $T_{ab}$ is an {\it unbiased estimator} for the 
function $\tau(\theta)$ if 
\begin{equation} 
T_{ab}\xi^{a}(\theta)\xi^{b}(\theta)\ =\ \tau(\theta)\ . 
\end{equation} 
For convenience, we define a mean-adjusted deviation 
operator $\tilde{T}_{ab} \equiv T_{ab}-\tau g_{ab}$. 
Note that ${\tilde T}_{ab}\xi^{a}\xi^{b}=0$, and that the 
variance of $T$ is given by ${\rm Var}_{\xi}[T] 
= \tilde{T}_{ab}\tilde{T}^{b}_{c}\xi^{a}\xi^{c}$. 
Since $T_{ab}\xi^{a}\xi^{b}=\tau$, we obtain 
$2T_{ab}\xi^{a}\dot{\xi}^{b}=\dot{\tau}$, hence 
$2{\tilde T}_{ab}\xi^{a}\dot{\xi}^{b}=\dot{\tau}$. 
Therefore, if we define $\eta_{b}= {\tilde T}_{ab} \xi^{a}$, 
we have $(\eta_{b}\dot{\xi}^{b})^{2} = \dot{\tau}^{2}/4$. 
Whence by use of the Cauchy-Schwartz inequality 
$(\eta^{a}\eta_{a}) (\dot{\xi}^{a} \dot{\xi}_{a}) \geq 
(\eta_{a}\dot{\xi}^{a})^{2}$, we are led to the following 
result. \par 

\begin{theo} 
${\rm (}$Cram\'er-Rao inequality${\rm )}$. Let $T$ 
be an unbiased estimator for a function $\tau(\theta)$ 
where $\theta$ parametrises a one-dimensional family of 
states $\xi^{a}(\theta)$ in ${\cal S}\in{\cal H}$. Then, the 
variance lower bound in the state $\xi^{a}$ is given by 
\begin{equation} 
{\rm Var}_{\xi}[T]\ \geq\ 
\frac{\dot{\tau}^{2}}{4\dot{\xi^{a}}\dot{\xi_{a}}} \ . 
\label{eq:CRI0} 
\end{equation} 
\end{theo} 

It is clear from the preceding argument that the CR lower bound is 
attained only if $\dot{\xi}^{a}=c\eta^{a}$ for some constant $c$, 
which by rescaling $\theta$ we can set to $1/2$ without loss of 
generality. Thus, for any curve $\xi^{a}(\theta)$ achieving the lower 
bound, we obtain the differential equation 
\begin{equation} 
\dot{\xi}^{a}\ =\ \frac{1}{2} 
\tilde{T}^{a}_{b}\xi^{b}\ , \label{eq:edif} 
\end{equation} 
The solution of (\ref{eq:edif}) is the {\it canonical exponential 
family} of distributions, given by the following elegant 
formula: 
\begin{equation} 
\xi^{a}(\theta)\ =\ 
\frac{\exp[\frac{1}{2}\theta T^{a}_{b}]q^{b}} 
{\sqrt{\exp[\theta T^{a}_{b}]q^{b}q_{a}}}\ , \label{eq:exp} 
\end{equation} 
where the normalised state vector $\xi^{a}(0) = 
q^{a}/(q_{b}q^{b})^{1/2}$ determines a prescribed initial 
distribution. Without loss of generality we can set $q_{a}q^{a}=1$. 
This expression leads us to an interesting geometrical 
interpretation of the exponential family. We consider the 
unit sphere $\cal{S}$ in $\cal{H}$, with the standard spherical 
geometry induced on it by $g_{ab}$. Let 
\begin{equation} 
\tau\ =\ \frac{T_{ab}\xi^{a}\xi^{b}}{g_{cd}\xi^{c}\xi^{d}} 
\label{eq:qftau} 
\end{equation} 
be a quadratic form defined on $\cal{S}$. Then $\cal{S}$ can be 
foliated by surfaces of constant $\tau$. Since according to 
(\ref{eq:edif}) the tangent vector $\dot{\xi}^{a}$ is parallel to 
the gradient of the function $\tau(\xi^{a})$, we conclude that: 

\begin{guess} 
The canonical exponential family of distributions 
$\xi^{a}(\theta)$, with initial distribution $q^{a}$, is given 
by the unique curve through the point $q^{a}$ that is everywhere 
orthogonal to the family of foliating $\tau$-surfaces. 
\label{prop:14} 
\end{guess} 

In particular, as we show in Proposition 6 below, the variance 
${\rm Var}_{\xi}[T]$ at the point $\xi^{a}$ is 
a quarter of the squared magnitude of the gradient 
of the surface through $\xi^{a}$, given by 
$\nabla_{a}\tau$. The Fisher information, on the other 
hand, is four times the squared magnitude of the tangent 
vector to the curve at $\xi^{a}$. Since the inner product 
of the tangent vector $\dot{\xi}^{a}$ and the normal 
vector $\nabla_{a}\tau$ is the derivative $\dot{\tau}$, 
it follows that 
${\rm Var}_{\xi}[T]\geq\dot{\tau}^{2}/{\cal G}$, the CR 
inequality. 

\begin{guess} 
Let $\nabla_{a}=\partial/\partial\xi^{a}$ denote the gradient 
operator in ${\cal H}$. Then the variance of an unbiased estimator 
$T_{ab}$ for a function $\tau$ in the state $\xi^{a}$ is given, 
on the sphere ${\cal S}$, by  
\begin{equation} 
{\rm Var}_{\xi}[T]\ =\ \frac{1}{4}\nabla_{a}\tau\nabla^{a}\tau\ . 
\label{eq:vart} 
\end{equation} 
\label{prop:21}
\end{guess} 

This can be verified as follows. By definition, we have a quadratic 
form (\ref{eq:qftau}) on ${\cal H}$. Then, by differentiation, we 
obtain 
\begin{eqnarray}  
\nabla_{a}\tau\ =\ \frac{2T_{ab}\xi^{b}}{\xi^{c}\xi_{c}} - 
\frac{2(T_{bc}\xi^{b}\xi^{c})\xi_{a}}{(\xi^{d}\xi_{d})^{2}}\ ,  
\end{eqnarray} 
from which it follows that 
\begin{equation}  
\frac{1}{4}\nabla_{a}\tau\nabla^{a}\tau\ =\ 
\frac{(T_{ac}T^{c}_{b}-\tau^{2}g_{ab})
\xi^{a}\xi^{b}}{(\xi^{c}\xi_{c})^{2}}\ . 
\label{eq:48} 
\end{equation} 
Since the variance of $T_{ab}$ is given by ${\rm Var}_{\xi}[T] = 
(T_{ac}T^{c}_{\ b} - \tau^{2}g_{ab})\xi^{a}\xi^{b}/\xi^{d}\xi_{d}$, 
equation (\ref{eq:vart}) follows at once after we restrict 
(\ref{eq:48}) to the sphere $\xi^{a}\xi_{a}=1$. \par 

In the case of the exponential family of distributions, the 
corresponding density function is given by 
$p(x|\theta) = q(x) \exp[ x\theta - \psi(\theta)]$, 
where $q(x)$ is the prescribed initial density, and the 
normalisation constant $\psi(\theta)$ is given by 
\begin{equation}  
\psi(\theta)\ =\ \ln \int_{-\infty}^{\infty} e^{x\theta} 
q(x)dx\ =\ \ln \left( \exp[\theta T^{b}_{a}]
q^{a}q_{b}\right) \ . 
\end{equation} 
\par 

It is interesting to note that the log-likelihood $l(x|\theta)$ 
for an exponential family has a natural geometric 
characterisation in 
${\cal H}$. Suppose we consider a multi-parameter exponential 
distribution given by 
\begin{equation} 
\xi^{a}(\theta)\ =\ \exp\left[ \frac{1}{2}\left( 
\sum_{j=1}^{r}\theta^{j} T^{a}_{(j)b} - 
\psi(\theta)\delta^{a}_{\ b}\right)\right] q^{b}\ . 
\end{equation} 
Our idea is to construct a random variable $l_{ab}$ in 
${\cal H}$ that represents the log-likelihood $l(x|\theta)$ 
for this family of distributions. We define the log-likelihood 
$\l^{a}_{\ b}$ associated with an exponential family of 
distributions by the symmetric operator 
\begin{equation} 
l^{a}_{\ b}(\theta)\ =\ \sum_{j=1}^{r}\theta^{j} 
T^{a}_{(j)b} - \psi(\theta)\delta^{a}_{\ b}\ . 
\end{equation} 
Note that the expectation of $l_{ab}$ gives the Shannon 
entropy, that is, 
\begin{equation} 
S_{\xi}(\theta)\ =\ l_{ab}\xi^{a}\xi^{b}\ =\ \sum_{j=1}^{r} 
\theta^{j}\tau_{(j)}(\theta) - \psi(\theta)\ . 
\end{equation} 
The second expression is the familiar one for the Legendre 
transformation that relates the entropy $S(\theta)$ to the 
normalisation constant $\psi(\theta)$. In the case of a 
one-parameter family of exponential distributions, the gradient 
$\nabla_{a}\tau$ can be written $\frac{1}{2}\nabla_{a}\tau = 
\dot{l}_{ab}\xi^{b}$. In the multi-parameter case this becomes 
$\frac{1}{2}\nabla_{a}\tau_{(j)} = \xi^{b}\partial_{j}l_{ab}$, 
which leads to the following formula for the Fisher 
information: \par 

\begin{guess} 
The Fisher information matrix ${\cal G}_{ij}$ can be expressed 
in terms of the log-likelihood $l^{a}_{\ b}$ by the formula 
${\cal G}_{ij} = \partial_{i}l_{ac}\partial_{j}l^{c}_{\ b} 
\xi^{a}\xi^{b}$. \label{prop:15} 
\end{guess} 
Thus in the case of an exponential family of distributions 
we find the Fisher-Rao metric is given by the covariance matrix 
of the estimators $T_{(i)}$: 
\begin{equation} 
{\cal G}_{ij}\ =\ (T_{(i)ac}-\partial_{i}\psi\delta_{ac}) 
(T^{c}_{(j)b}-\partial_{j}\psi\delta^{c}_{\ b})\xi^{a}\xi^{b} 
\ \equiv\ E_{\xi}[{\tilde T}_{(i)}{\tilde T}_{(j)}]\ . 
\end{equation} 
\par 

\section{Generalised Bhattacharyya Bounds} 

We have observed that the exponential family is the only family 
capable of achieving the variance lower bound, providing we choose 
the right function $\tau(\theta)$ of the parameter to estimate. For 
other families of distributions, the variance exceeds the lower 
bound. In order to obtain sharper bounds in the general situation, 
we consider the possibility of establishing higher-order 
corrections to the CR lower bound. Our approach is related to 
that of Bhattacharyya (1946, 1947, 1948). However, in a Hilbert 
space context, we are led along a different route from 
Bhattacharyya's original considerations, since in his approach 
the likelihood function $p(x|\theta)$ plays a crucial role. First, 
we shall formulate a new, Bhattacharyya-style derivation of the 
CR inequality. We note that if $T_{ab}$ is an unbiased estimator 
for the function $\tau(\theta)$, then so is 
$R_{ab} = T_{ab}+\lambda\xi_{(a}\dot{\xi}_{b)}$, for an 
arbitrary constant $\lambda$. We choose the value of $\lambda$ that 
minimises the variance of $R_{ab}$. This implies 
$\lambda=-\dot{\tau}/2\dot{\xi}^{a}\dot{\xi}_{a}$, and 
hence 
\begin{equation}  
{\rm min}({\rm Var}_{\xi}[R])\ =\ {\rm Var}_{\xi}[T] 
- \frac{\dot{\tau}^{2}}{4\dot{\xi}^{a}\dot{\xi}_{a}}\ . 
\end{equation} 
Since ${\rm Var}_{\xi}[R]\geq0$, we are immediately 
led back to the CR inequality (\ref{eq:CRI0}).\par 

Now we try to improve on this by incorporating terms with 
higher-order derivatives. Let us denote the $r$-th derivative 
of $\xi^{a}$ with respect to the parameter $\theta$ by 
$\xi^{(r)a} = d^{r}\xi^{a} / d\theta^{r}$. We write 
$\hat{\xi}^{(r)a}$ for the projection of $\xi^{(r)a}$ orthogonal 
to $\xi^{a}$ and to all the lower order derivatives, so 
$\hat{\xi}^{(r)}_{a}\xi^{a}=0$ and $\hat{\xi}^{(r)}_{a}\xi^{(s)a}=0$ 
for $s<r$. If $T_{ab}$ is an unbiased estimator for $\tau(\theta)$, 
so is the symmetric tensor $R_{ab}$ defined by 
\begin{equation} 
R_{ab}\ =\ T_{ab} + 
\sum_{r}\lambda_{r}\xi_{(a}\hat{\xi}^{(r)}_{b)}
\end{equation} 
for arbitrary constants $\lambda_{r}$. We only consider values of 
$r$ such that $\hat{\xi}^{(r)}_{a}\neq 0$, assuming that the relevant 
derivatives exist and are linearly independent. A straightforward 
calculation leads us to the values of $\lambda_{r}$ minimising the 
variance of $R$, and we obtain 
\begin{equation}  
{\rm min}({\rm Var}_{\xi}[R])\ =\ {\rm Var}_{\xi}[T] - 
\sum_{r}\frac{(T_{ab}\xi^{a}\hat{\xi}^{(r)b})^{2}}
{g_{ab}\hat{\xi}^{(r)a}\hat{\xi}^{(r)b}}\ . 
\end{equation} 
Since ${\rm Var}_{\xi}[R]$ is nonnegative, we thus 
deduce the following {\it generalised Bhattacharyya bounds} 
for the variance of the estimator: 
\begin{equation} 
{\rm Var}_{\xi}[T]\ \geq\ 
\sum_{r} \frac{(T_{ab}\xi^{a}\hat{\xi}^{(r)b})^{2}}
{g_{ab}\hat{\xi}^{(r)a}\hat{\xi}^{(r)b}}\ . \label{eq:Bhb} 
\end{equation} 

This derivation is `historical' in flavour in the sense that it 
parallels certain aspects of the original argument of Bhattacharyya. 
However, Proposition \ref{prop:21} allows us to reexpress 
(\ref{eq:Bhb}) in the form of a simple geometric inequality. That 
is, given the gradient vector $\nabla_{a}\tau$ in ${\cal H}$, the 
squared length of this vector is not less than the sum of the 
squares of any of its orthogonal components with respect to a 
suitable basis. To this end, we consider the vectors based on the 
state $\xi^{a}$ and its higher order derivatives, and form the 
orthonormal vectors given by 
$\hat{\xi}^{(r)a}/(\hat{\xi}^{(r)b}\hat{\xi}^{(r)}_{b})^{1/2}$. It 
follows from the basic relation (\ref{eq:vart}) deduced in 
Proposition 6 that: 

\begin{guess} 
The generalised variance lower bounds for an unbiased estimator 
$T$ of a function $\tau$ can be expressed in the form: 
\begin{equation} 
{\rm Var}_{\xi}[T]\ \geq\ \frac{1}{4} \sum_{r}
\frac{(\hat{\xi}^{(r)a}\nabla_{a}\tau)^{2}}
{\hat{\xi}^{(r)b}\hat{\xi}^{(r)}_{b}}\ . \label{eq:gbb3} 
\end{equation} 
\label{prop:Bhat} 
\end{guess} 

Clearly for $r=1$ we recover the CR inequality. Unlike the classical 
Bhattacharyya bounds, the generalised bounds are not necessarily 
independent of the estimator $T$. In our applications to quantum 
mechanics, however, we shall indicate some important examples of 
higher-order bounds that are indeed independent of the specific 
choice of estimator. See Brody and Hughston (1996d) for a 
related example drawn from classical thermal physics where the bounds 
are also systematically independent of the estimator. \par 

We remark, incidentally, that the denominator terms in equation 
(\ref{eq:gbb3}) give rise to natural geometric invariants. 
For example, in the case $r=2$ we have 
\begin{equation} 
\hat{\xi}^{(2)a}\hat{\xi}^{(2)b}g_{ab}\ =\ 
(\dot{\xi}^{a}\dot{\xi}^{b}g_{ab})^{2} K^{2}_{\xi}\ , 
\end{equation} 
where $K_{\xi}^{2}$ is the curvature of the curve $\xi^{a}(\theta)$ 
in ${\cal S}$. In particular, we obtain 

\begin{lem} 
In the case of the canonical exponential family of distributions, 
specified in equation ${\rm (\ref{eq:exp})}$, the curvature of 
$\xi^{a}(\theta)$ is given by:  
\begin{equation} 
K_{\xi}^{2}\ =\ 
\frac{\langle {\tilde T}^{4}\rangle}
{\langle{\tilde T}^{2}\rangle^{2}} - 
\frac{\langle{\tilde T}^{3}\rangle^{2}}
{\langle{\tilde T}^{2}\rangle^{3}} - 1\ . \label{eq:statcurv} 
\end{equation}   
\label{lem:12} 
\end{lem} 

As a matter of interpretation we note that the first term in the 
right hand side of (\ref{eq:statcurv}) is the kurtosis (measure of 
sharpness) of the distribution, while the second term is the 
skewness (measure of asymmetry). A classical statistical inequality 
relating these quantities (cf. Stuart and Ord 1994) ensures 
that $K_{\xi}^{2}\geq0$. In the case of the exponential family we 
have $\hat{\xi}^{(2)a}\nabla_{a}\tau=0$, i.e., the `acceleration 
vector' $\hat{\xi}^{(2)a}$ lies in the tangent space of the 
surfaces generated by constant values of the estimator function 
$\tau(\theta)$. \par 
 
\section{Multiple Parameters} 

The geometrical constructions so far considered are based mainly 
upon one-parameter families of distributions. However, for 
completeness here we sketch some useful results applicable to 
multi-parameter distributions. First, consider the case where we 
estimate a single function $\tau(\theta)$ depending upon several 
parameters $\theta^{i}$. A straightforward argument shows that the 
CR inequality then takes the form 
\begin{equation} 
{\rm Var}_{\xi}[T]\ \geq\ \sum_{ij} 
{\cal G}^{ij}\tau_{i}\tau_{j}\ , \label{eq:mpcr} 
\end{equation} 
where $\tau_{i}=\partial_{i}\tau(\theta)$ and 
${\cal G}^{ij}$ 
is the inverse of the Fisher information matrix. 
In a more general situation, we might have several estimators 
$T_{(\alpha) ab}$ $(\alpha=1, \cdots, n)$ labelled by an index 
$\alpha$, 
with $T_{(\alpha) ab}\xi^{a}\xi^{b} = \tau_{\alpha}(\theta)$. 
For an arbitrary set of constants $\Lambda_{\alpha}$, we form the 
sums $T_{ab} = \sum_{\alpha}\Lambda_{\alpha}T_{(\alpha) ab}$ and 
$\tau(\theta) = \sum_{\alpha}\Lambda_{\alpha}\tau_{\alpha}$.  
It follows that the CR inequality (\ref{eq:mpcr}) holds 
for the summed expressions $T_{ab}$ and $\tau(\theta)$. 
However, since $\Lambda_{\alpha}$ is constant, 
the variance of $T$ can be written 
\begin{equation} 
{\rm Var}_{\xi}[T]\ =\ \sum_{\alpha\beta}
C_{\alpha\beta}\Lambda_{\alpha}\Lambda_{\beta}\ , 
\end{equation} 
where $C_{\alpha\beta} = {\rm Cov}_{\xi}[T_{(\alpha)},T_{(\beta)}]$ 
is the covariance matrix for the estimators $T_{(\alpha)}$. 
Therefore, the CR inequality can be rewritten in the form  
\begin{equation}  
\sum_{\alpha\beta}{\rm Cov}_{\xi}[T_{(\alpha)},T_{(\beta)}] 
\Lambda_{\alpha}\Lambda_{\beta} -\sum_{\alpha\beta} 
{\cal G}^{ij}\partial_{i}\tau_{\alpha} 
\partial_{j}\tau_{\beta}  
\Lambda_{\alpha}\Lambda_{\beta}\ \geq\ 0\ . 
\end{equation} 
Since this holds for arbitrary values of 
$\Lambda_{\alpha}$, we obtain the following matrix 
inequality for the covariance lower bound. \par 

\begin{guess} 
Let $T_{(\alpha)}$ $(\alpha=1, 2, \cdots, r)$ 
be unbiased estimators for the functions 
$\tau_{\alpha}(\theta)$. The lower bound for the 
covariance matrix is given by 
\begin{equation} 
{\rm Cov}_{\xi}[T_{(\alpha)},T_{(\beta)}]\ \geq\ 
{\cal G}^{ij}\partial_{i}\tau_{\alpha} 
\partial_{j}\tau_{\beta} \ . 
\end{equation} 
\label{prop:16} 
\end{guess} 
This equation is to be interpreted in the sense of saying that the 
difference between the left and right hand sides is nonnegative 
definite. \par 

\section{From Classical to Quantum Theory} 

In the foregoing material, we have reformulated various aspects of 
parametric statistical inference in terms of the geometry of a real 
Hilbert space. In particular, the abstract index notation has enabled 
us very efficiently to obtain results relating to statistical 
curvatures and variance lower bounds. One of the main reasons we 
are interested in formulating statistical estimation theory in a 
Hilbert space framework is on account of the connection with 
quantum mechanics, which becomes more direct when pursued in this 
manner, thus enabling us in many respects to unify our view of 
classical and quantum statistical estimation. \par 

The fact that in our approach to classical statistical estimation 
the geometry in question is a Hilbert space geometry is a result 
that physicists may find surprising. This is because the general view 
in physics is that the Hilbert space structure associated with the 
space of states in nature is special to quantum theory, and has no 
analogue in classical probability theory and statistics. We have 
seen, however, that a number of structures already present in the 
classical theory are highly analogous to associated quantum mechanical 
structures; but the correspondence is only readily apparent when 
the classical theory is reformulated in the appropriate geometrical 
framework. A key point is that if we supplement the real Hilbert 
space ${\cal H}$ with a {\it compatible complex structure}, 
then this paves the way for a natural attack on problems of quantum 
statistical inference, and it becomes possible to see more clearly 
which aspects of statistical inference are universal, and which are 
particular to the classical or quantum domain. \par 

Indeed, there are a number of distinct geometrical formulations of 
classical statistical theory, corresponding, for example, to the 
various $\alpha$-embeddings of Amari (see, e.g., Amari 1985 or 
Murray and Rice 1993), but one among these is singled out on account 
of its close relation to quantum theory: the geometry of square-root 
density functions. This geometry is special because of the way it 
singles out the Levi-Civita connection on statistical submanifolds, 
as indicated in Proposition 2 above. In this way we are led to 
consider classical statistics in the language of real Hilbert space 
geometry, as indicated in the previous sections. The real Hilbert 
space formulation of standard quantum theory, on the other hand, 
is in itself a fairly standard construction now, though perhaps not 
as well known as it should be, and in the next section we shall 
develop some of the formalism necessary for working in this 
framework. \par 

The specific point of originality in our approach is to make the 
link between the natural real Hilbert space arising on the one hand 
in connection with the classical theory of statistical inference, 
with the natural real Hilbert space arising on the other hand in 
connection with standard quantum theory. Once this identification 
has been made, then a number of interesting results can be seen to 
follow, which are explored in some detail here. In particular, the 
theory of classical statistical estimation can be extended directly 
to the quantum mechanical situation, and we are able to show how the 
Cram\'er-Rao inequality associated with a pair of canonically 
conjugate physical 
variables can be interpreted as the corresponding Heisenberg relation 
in the quantum mechanical context. This ties in neatly with the 
important line of investigation in quantum statistical estimation 
initiated by Helstrom (1969), Holevo (1973, 1979), and others (e.g., 
Yuen, Kennedy, and Lax 1975), about which we shall have more to say 
shortly. One of the most exciting results emerging as a by-product 
of our approach is the development of a series of `improved' 
Heisenberg relations, formulated in some detail in the later 
sections. \par 

\section{Geometry of Quantum States} 

Now we turn to quantum geometry. Our goal in this section is to 
formulate standard quantum theory in a geometrical language that 
brings out more clearly its relation to the statistical geometry 
which we have developed in sections 2 to 6. We start with our 
formulation of classical inference, based on a real Hilbert space 
geometry, upon which we will now impose additional structure. Thus 
instead of `completely reformulating everything' from scratch to 
develop a quantum statistical theory, as has conventionally been 
done, we shall essentially accept the classical theory, but 
`enrich' it with some extra structure. The essential additional 
ingredient that we must introduce on our real Hilbert space 
$\cal{H}$, in order to study quantum mechanical systems, is, more 
specifically, a {\it compatible complex structure}. A complex 
structure on ${\cal H}$ is given by a tensor $J^{b}_{\ a}$ satisfying 
\begin{equation}  
J^{b}_{\ a}J^{c}_{\ b} \ =\ - \delta^{c}_{\ a}\ . \label{eq:comp} 
\end{equation} 
Given this structure we then say a symmetric operator $X_{ab}$ is 
{\it Hermitian} if it satisfies the relation 
\begin{equation}  
J^{a}_{\ c}J^{b}_{\ d}X_{ab} \ =\ X_{cd}\ . \label{eq:herm} 
\end{equation} 
An alternative way to express the Hermitian condition is 
$J^{a}_{\ c}X^{c}_{\ b} = X^{a}_{\ c}J^{c}_{\ b}$, which states that 
$J^{a}_{\ b}$ and $X^{a}_{\ b}$ commute. This follows from the 
complex structure identity (\ref{eq:comp}) and the Hermiticity 
condition (\ref{eq:herm}). We require that the complex structure be 
compatible with the Hilbert space structure by insisting that the 
metric $g_{ab}$ is Hermitian. As a consequence we have 
$J^{a}_{\ c}J^{b}_{\ d}g_{ab}  = g_{cd}$,  
which is to be viewed as a fundamental relationship holding 
between $J^{a}_{\ b}$ and $g_{ab}$. \par 

In order to proceed further it will be useful to make a comparison 
of the index notation being used here with the conventional Dirac 
notation. In the `real' approach to quantum theory, state vectors are 
represented by elements of a real Hilbert space ${\cal H}$. We find 
that if $\xi^{a}$ and $\eta^{a}$ are two real Hilbert space vectors, 
then their Dirac product is given by the following complex expression: 
\begin{equation} 
\langle\eta|\xi\rangle\ =\ \frac{1}{2} \eta^{a}(g_{ab} 
- ig_{ac}J^{c}_{\ b})\xi^{b}\ . \label{eq:dp} 
\end{equation}  
The Hermitian property of $g_{ab}$ implies that the tensor 
$\Omega_{ab} = g_{ac}J^{c}_{\ b}$ is automatically antisymmetric 
and invertible, i.e., a {\it symplectic structure}, which also 
satisfies the Hermitian condition in the 
sense that $J^{a}_{\ c}J^{b}_{\ d}\Omega_{ab} = \Omega_{cd}$.   
Since the symplectic structure $\Omega_{ab}$ is antisymmetric, it 
follows then that the Dirac norm agrees with the real Hilbertian norm 
(apart from the factor of two): 
\begin{equation} 
\langle\xi|\xi\rangle\ =\ 
\frac{1}{2}g_{ab}\xi^{a}\xi^{b}\ . 
\end{equation} 
\par 

A real Hilbert space vector $\xi^{a}$ can be decomposed into 
complex `positive' and `negative' parts, relative to 
the specified complex structure, according to the scheme 
$\xi^{a} = \xi^{a}_{+} + \xi^{a}_{-}$, where 
\begin{equation}  
\xi^{a}_{\pm}\ \equiv\ \frac{1}{2} 
(\xi^{a}\mp iJ^{a}_{\ b}\xi^{b})\ . 
\end{equation} 
In the case of relativistic fields, this decomposition corresponds 
to splitting the fields into positive and negative frequency 
parts, so occasionally we refer to $\xi^{a}_{+}$ and $\xi^{a}_{-}$ as 
the `positive frequency' and `negative frequency' parts of $\xi^{a}$. 
Note that $\xi^{a}_{+}$ and $\xi^{a}_{-}$ are complex `eigenstates' 
of the $J^{a}_{\ b}$ operator, in the sense that 
$J^{a}_{\ b}\xi^{b}_{\pm} = \pm i \xi^{a}_{\pm}$. 
As a consequence, the Hermitian condition (\ref{eq:herm}) implies 
that two vectors of the same `type' (e.g., a pair of positive 
vectors) are necessarily orthogonal with respect to the metric 
$g_{ab}$. In other words, we have 
$g_{ab}\xi^{a}_{\pm}\eta^{b}_{\pm}=0$ for any two positive (negative) 
vectors $\xi^{a}_{\pm}$ and $\eta^{a}_{\pm}$. \par 

For certain purposes it is useful to introduce Greek 
indices to denote positive and negative parts, by writing 
$\xi^{a} = (\xi^{\alpha}, \bar{\xi}_{\alpha})$, 
where $\bar{\xi}_{\alpha}$ is the complex conjugate of 
$\xi^{\alpha}$. Then, we can identify $\xi^{\alpha}$ with the Dirac 
`ket' vector $|\xi\ra$, and $\bar{\xi}_{\alpha}$ with the complex 
conjugate Dirac `bra' vector $\la\bar{\xi}|$, and write 
$\xi^{a} = (|\xi\rangle, \langle\bar{\xi}|)$. 
To be more specific, a typical element in the complex Hilbert 
space is denoted $\psi^{\alpha}$, or equivalently $|\psi\rangle$ 
in the conventional 
Dirac notation, and an element in the dual space is denoted 
$\varphi_{\alpha}=\langle\varphi|$. Hence, their inner product 
is written $\varphi_{\alpha}\psi^{\alpha} = 
\langle\varphi|\psi\rangle$. The complex conjugate of the vector 
$\psi^{\alpha}$ is 
$\bar{\psi}_{\alpha}=\langle\bar{\psi}|$, and its norm is then 
given by $\bar{\psi}_{\alpha}\psi^{\alpha} = 
\langle\bar{\psi}|\psi\rangle$. 
If we denote the splitting of a real Hilbert space ${\cal H}$ 
into positive and negative eigenspaces by 
${\cal H} = {\cal H}^{+} \oplus {\cal H}^{-}$, 
then an `operator' in quantum mechanics can be regarded as a 
linear map $T^{\alpha}_{\ \beta}$ from a domain in ${\cal H}^{+}$ 
to ${\cal H}^{+}$, given, e.g., by 
$T^{\alpha}_{\ \beta}\psi^{\beta} = \eta^{\alpha}$, for which 
the corresponding complex conjugate operator is 
$\overline{T^{\alpha}_{\ \beta}}=\overline{T}_{\alpha}^{\ \beta}$. 
Thus, if $T$ is Hermitian, we have 
$T^{\alpha}_{\ \beta}=\overline{T}^{\alpha}_{\ \beta}$, and it 
follows that the expectation $\langle T\rangle$ of $T$ in the 
state $\psi$ is given by 
\begin{equation} 
\frac{\langle\bar{\psi}|T|\psi\rangle}
{\langle\bar{\psi}|\psi\rangle}\ =\ 
\frac{\bar{\psi}_{\alpha}T^{\alpha}_{\ \beta}\psi^{\beta}}
{\psi^{\alpha}\bar{\psi}_{\alpha}} \ , 
\end{equation} 
and the variance of $T$ is 
\begin{equation} 
{\rm Var}_{\psi}[T]\ =\ \frac{\bar{\psi}_{\alpha}
{\tilde T}^{\alpha}_{\ \gamma}{\tilde T}^{\gamma}_{\ \beta}  
\psi^{\beta}}{\psi^{\alpha}\bar{\psi}_{\alpha}} \ , 
\end{equation} 
where ${\tilde T}^{\alpha}_{\ \beta} = T^{\alpha}_{\ \beta} 
- \langle T\rangle\delta^{\alpha}_{\ \beta}$. Note that 
$\langle {\tilde T}\rangle = 0$. \par 

Now let us say more about the Hermitian condition. Having the 
decomposition $\xi^{a} = (\xi^{\alpha},\bar{\xi}_{\alpha})$ in mind, 
we can represent a given second rank tensor $T_{ab}$ (not 
necessarily symmetric, Hermitian, or real) in matrix form by writing 
\begin{equation}  
T_{ab}\ =\ \left( \begin{array}{cc} 
A_{\alpha\beta} & B_{\alpha}^{\ \beta} \\ 
C^{\alpha}_{\ \beta} & D^{\alpha\beta} \end{array} 
\right)\ .  
\end{equation} 
Similarly, the complex structure $J^{a}_{\ b}$ can be written 
\begin{equation} 
J^{a}_{\ b}\ =\ \left( \begin{array}{cc} 
i\delta^{\alpha}_{\ \beta} & 0 \\ 
0 & -i\delta_{\alpha}^{\ \beta} \end{array} \right)  \ , 
\end{equation} 
Thus for the action of the complex structure tensor we find 
$\zeta^{a} \equiv J^{a}_{\ b}\xi^{b} = 
(i\xi^{\alpha}, -i\bar{\xi}_{\alpha})$. 
In other words, the effect of $J^{a}_{\ b}$ is to multiply the 
`ket' part of the given state by $i$, and the `bra' part by $-i$. 
Moreover, it can be verified that 
\begin{equation}  
J^{c}_{\ a}J^{d}_{\ b}T_{cd}\ =\ \left( \begin{array}{cc}  
- A_{\alpha\beta} & B_{\alpha}^{\ \beta} \\ 
C^{\alpha}_{\ \beta} & -D^{\alpha\beta} \end{array} 
\right)\ .  
\end{equation} 
Therefore, the requirement that the tensor $T_{ab}$ should be 
symmetric implies $A_{\alpha\beta}=A_{(\alpha\beta)}$, 
$D^{\alpha\beta}=D^{(\alpha\beta)}$, and $B_{\alpha}^{\ \beta} 
= C^{\beta}_{\ \alpha}$. The Hermitian condition then implies 
$A_{\alpha\beta}=0$ and $D^{\alpha\beta}=0$, and the reality 
condition implies $B_{\alpha}^{\ \beta} = 
{\overline C}^{\beta}_{\ \alpha}$. A symmetric, real Hermitian 
tensor $T_{ab}$ can be represented in matrix form by writing 
\begin{equation}  
T_{ab}\ =\ \left( \begin{array}{cc} 
0 & T^{\alpha}_{\ \beta} \\ 
T_{\alpha}^{\ \beta} & 0 \end{array} \right)\ , 
\end{equation}  
where ${\overline T}^{\alpha}_{\ \beta}=T_{\beta}^{\ \alpha}$. It 
follows that the quadratic form $T_{ab}\xi^{a}
\eta^{b}$ for a Hermitian tensor $T_{ab}$ is given by  
$T_{ab}\xi^{a}\eta^{b} = T^{\alpha}_{\ \beta}
\bar{\xi}_{\alpha}\eta^{\beta} + T_{\alpha}^{\ \beta}
\xi^{\alpha}\bar{\eta}_{\beta}$. 
In the special case of the metric $g_{ab}$ we have 
\begin{equation} 
g_{ab}\ =\ \left( \begin{array}{cc} 
0 & \delta^{\alpha}_{\ \beta} \\ 
\delta_{\alpha}^{\ \beta} & 0 \end{array} \right)\ , 
\end{equation} 
from which it follows that $g_{ab}\xi^{a}\eta^{b} = 
\xi^{\alpha}\bar{\eta}_{\alpha} + \bar{\xi}_{\alpha}\eta^{\alpha}$. 
Also, for the symplectic structure $\Omega_{ab}$ we obtain 
\begin{equation} 
\Omega_{ab}\ =\ \left( \begin{array}{cc} 
0 & i\delta^{\alpha}_{\ \beta} \\ 
-i\delta_{\alpha}^{\ \beta} & 0 \end{array} \right)\ , 
\end{equation} 
so that $\Omega_{ab}\xi^{a}\eta^{b} = 
i\bar{\xi}_{\alpha}\eta^{\alpha} - 
i\xi^{\alpha}\bar{\eta}_{\alpha}$. 
Clearly, we then have 
\begin{equation} 
\xi^{\alpha}\bar{\eta}_{\alpha}\ =\ \frac{1}{2} \left( 
g_{ab} + i \Omega_{ab} \right) \xi^{a}\eta^{b}\ , 
\end{equation} 
which is consistent with equation (\ref{eq:dp}), if we bear 
in mind that $\Omega_{ab}$ is antisymmetric. With these 
relations at hand, the reformulation in `real' terms of the 
standard `complex' formalism of quantum theory can be pursued 
in a straightforward, systematic way. It is worth bearing in 
mind that in formulating quantum theory in real terms in this way 
we have not altered the results or physical content 
of the theory. These remain unchanged. Our purpose, rather, is to 
highlight in this way certain geometrical and probabilistic 
aspects of ordinary quantum mechanics that otherwise may not be 
obvious. 
For further details of the `real' approach to complex Hilbert space 
geometry and its significance in quantum mechanics, see for example 
Ashtekar and Schilling (1995), Field (1996), Geroch (1971a,b), 
Gibbons (1992), Gibbons and Pohle (1993), Kibble (1978,1979), 
Schilling (1996), Segal (1947), and Wald (1976,1994). \par 

\section{Real Hilbert Space Dynamics} 

In this section we take the discussion further by consideration of 
the quantum mechanical commutation relations, as seen from a `real' 
Hilbert space point of view. This then leads us to a natural `real' 
formulation of the Schr\"odinger equation. \par

If $X_{ab}$ and $Y_{ab}$ are a pair of symmetric operators, then 
their `skew product' defined by the expression 
$X_{a}^{\ c}Y_{cb} - X_{b}^{\ c}Y_{ca}$ is an antisymmetric tensor, 
and thus itself cannot represent a random variable. Nevertheless, 
in the case of Hermitian operators, there is a natural isomorphism 
between the space of symmetric tensors 
$X_{ab}$ satisfying $J^{c}_{\ a}J^{d}_{\ b}X_{cd}=X_{ab}$ 
and antisymmetric tensors $\Lambda_{ab}$ satisfying 
$J^{c}_{\ a}J^{d}_{\ b}\Lambda_{cd}=\Lambda_{ab}$, and the map 
in question is given by contraction with $J^{a}_{\ b}$. This 
follows from the fact that if $X_{ab}$ is symmetric and 
Hermitian, then $\Lambda_{ab}=X_{ac}J^{c}_{\ b}$ is automatically 
antisymmetric and Hermitian. 
Conversely, if $\Lambda_{ab}$ is antisymmetric and Hermitian, then 
$\Lambda_{ac}J^{c}_{\ b}$ is automatically symmetric and Hermitian. 
Thus to 
form the commutator of two symmetric Hermitian operators first 
we take their skew product, which then we multiply by the 
complex structure tensor to give us a symmetric Hermitian 
operator. After some rearrangement of terms, these results can 
be summarised as follows: \par 

\begin{lem} 
The commutator $Z = i[X,Y]$ of a pair of symmetric, Hermitian 
operators $X$ and $Y$ is given by the symmetric Hermitian 
operator $Z_{ab} = (X_{ac}Y_{bd} - Y_{ac}X_{bd})\Omega^{cd}$. 
\label{lem:21} 
\end{lem} 

Note that the symplectic structure $\Omega_{ab}$ (or equivalently, 
the complex structure) is playing the role of `$i$' in the relation 
$Z=i[X,Y]$ so as to give us a real, symmetric, Hermitian tensor 
$Z_{ab}$. \par 

The anticommutator $W=\{ X,Y\}$ between two symmetric operators 
$X_{ab}$ and $Y_{ab}$ is defined by $W_{ab}=2X_{(a}^{\ c}Y_{b)c}$. 
This is a more `primitive' operation on the space of symmetric 
tensors since it does not require introduction of a complex 
structure. The basic operator identity 
\begin{equation}  
\{\{A,B\},C\} - \{A,\{B,C\}\}\ =\ [B,[A,C]] \label{eq:compat} 
\end{equation} 
shows that even in the absence of a Hermitian structure the 
incompatibility between a pair of random variables can be 
expressed in terms of the nonassociativity of the symmetric 
product. In other words, we say two random variables $A$ and 
$C$ are compatible iff the left-hand side of (\ref{eq:compat}) 
vanishes for any choice of $B$. \par 

Now, suppose the Hamiltonian of a quantum mechanical system is 
represented by a symmetric Hermitian operator $H_{ab}$. In fact, 
we need $H_{ab}$ to be self-adjoint (a stronger condition), but 
this need not concern us for the moment. Then for the Schr\"odinger 
equation we have: 
\begin{equation}  
\dot{\xi}^{a}\ =\ 
J^{a}_{\ b}H^{b}_{\ c}\xi^{c}\ . \label{eq:schro}
\end{equation} 
Note that again the role of the usual `$i$' factor is played by the 
complex structure tensor. Expressing this relation in terms of 
positive and negative parts, we then recover the conventional form 
of the Schr\"odinger equation $i\dot{\xi}^{a}_{+} = 
H^{a}_{\ b}\xi^{b}_{+}$, together with its complex conjugate. In 
Dirac's notation this is of course $i\partial_{t}|\xi\ra = H|\xi\ra$. 
As a consequence of (\ref{eq:schro}) it follows at once that 
$\dot{\xi}^{a}\xi_{a}=0$. This is due to the Hermitian relation 
which says that $J_{ac}H^{c}_{\ b}$ is antisymmetric. Thus, as 
expected, the Schr\"odinger equation respects the normalisation 
$g_{ab}\xi^{a}\xi^{b}=1$. \par 

Having formulated the conventional quantum dynamics in terms of 
real Hilbert space, we are in a position to make an interesting link 
with statistical considerations. To begin, we note that the usual 
phase freedom in quantum mechanics can be incorporated by modifying 
the Hamiltonian according to the prescription  
$H^{b}_{\ a}  \rightarrow \tilde{H}^{b}_{\ a} = H^{b}_{\ a} + 
\varphi\delta^{b}_{\ a}$. 
We can take advantage of this freedom by consideration of the 
following result. 

\begin{guess} 
There is a unique choice of phase such that the tangent vector 
$\dot{\xi}^{a}$ of the dynamical trajectory is everywhere 
orthogonal to the direction 
$\zeta^{a}=J^{a}_{\ b}\xi^{b}$. This choice of $\varphi$ minimises 
the Fisher information $4g_{ab}\dot{\xi}^{a}\dot{\xi}^{b}$. 
\label{lem:22} 
\end{guess} 

In fact, the relevant phase factor is given 
by $\varphi = -H_{ab}\xi^{a}\xi^{b}/\xi^{c}\xi_{c}$. 
Physically, this choice of phase fixing implies an adjustment 
of the mean of the Hamiltonian. Clearly, we have 
${\tilde H}_{ab}\xi^{a}\xi^{b}=0$, and it is not difficult 
to see that the same choice of $\varphi$ minimises 
$4g_{ab}\dot{\xi}^{a}\dot{\xi}^{b}$. In fact, for general 
$\varphi$ we have 
$g_{ab}\dot{\xi}^{a}\dot{\xi}^{b} = 
g_{ab}H^{a}_{\ c}H^{b}_{\ d}\xi^{c}\xi^{d} + 
2\varphi H_{ab}\xi^{a}\xi^{b} + \varphi^{2}g_{ab}\xi^{a}\xi^{b}$, 
from which it follows at once that 
$g_{ab}\dot{\xi}^{a}\dot{\xi}^{b}$ is minimised for the choice of 
$\varphi$ indicated. This result will be used 
extensively in our work on quantum estimation. With this choice 
of phase the modified Schr\"odinger equation reads 
\begin{equation} 
\dot{\xi}^{a}\ =\ 
J^{a}_{\ b}{\tilde H}^{b}_{\ c}\xi^{c}\ , \label{eq:sch2} 
\end{equation} 
where 
\begin{equation} 
{\tilde H}_{ab}\ =\ H_{ab} - \left( \frac{H_{cd}\xi^{c}\xi^{d}}
{\xi^{e}\xi_{e}}\right) g_{ab} 
\end{equation}  
represents now the deviation of the Hamiltonian from its mean, 
in accordance with the notation introduced earlier. Note that 
for the state defined by $\zeta^{a}=J^{a}_{\ b}\xi^{b}$, the 
dynamical equation becomes $\dot{\zeta}^{a} = 
J^{a}_{\ b}H^{b}_{\ c}\zeta^{c}$, since $J^{a}_{\ b}$ commutes with 
$H^{a}_{\ b}$. Thus, $\zeta^{a}$ also satisfies the Schr\"odinger 
equation. We can think of the complex projective space (in general 
infinite dimensional) formed by projectivising the `positive' 
Hilbert space ${\cal H}^{+}$ as being the `true' space of pure 
states. Then the essence of Proposition \ref{lem:22} is that there 
is a unique `lift' from this projective space $P({\cal H}^{+})$ to 
the real Hilbert space ${\cal H}$ such that the tangent vector 
$\dot{\xi}^{a}$ is everywhere orthogonal to both $\xi^{a}$ and 
$\zeta^{a}$. \par 

As a matter of interpretation we make the following observation 
regarding the `modified' Schr\"odinger equation. In the standard 
treatment of quantum mechanics one is taught that the time 
independent Schr\"odinger equation is given by 
$H|\xi\rangle=E|\xi\rangle$, whereas the time dependent case can 
be written by use of the correspondence principle 
$E\leftrightarrow i\partial_{t}$. Although generally accepted, 
the basis of this correspondence has to be regarded as somewhat 
mysterious, and to that extent also unsatisfactory. Now, in the 
modified Schr\"odinger equation we have $i\partial_{t} 
|\xi\rangle = (H-\langle H\rangle)|\xi\rangle$. Hence if the 
state is time independent, we recover the usual time 
independent equation $(H-E)|\xi\rangle=0$. In this way, we do 
not have to specify which representation of the canonical 
commutation relations we work with. While in general terms the 
theory is independent of the specific choice of phase, it seems 
that there is a unique choice of phase that makes everything 
fit in well from a physical point of view, and interestingly 
we are led to the same result from purely statistical 
considerations. \par 

Now suppose $B_{ab}$ is a symmetric Hermitian operator and we write 
$B(t) := E_{\xi(t)}[B]$ for the expectation 
\begin{equation} 
E_{\xi(t)}[B]\ =\ \frac{B_{ab}\xi^{a}(t)\xi^{b}(t)}
{g_{cd}\xi^{c}(t)\xi^{d}(t)}\ , 
\end{equation} 
where $\xi^{a}(t)$ satisfies the modified Schr\"odinger equation 
(\ref{eq:sch2}), or equivalently, $\xi^{a}(t)=\exp[tJ^{a}_{\ b} 
{\tilde H}^{b}_{\ c}]\xi^{c}(0)$, where $\xi^{a}(0)$ is the state 
vector at $t=0$. Thus $B(t)$ represents the expectation of $B_{ab}$ 
along the given trajectory. It follows that 
\begin{equation} 
\frac{dB(t)}{dt}\ =\ \frac{C_{ab}\xi^{a}\xi^{b}}
{g_{cd}\xi^{c}\xi^{d}} 
\end{equation} 
along $\xi^{a}(t)$, where $C_{ab}=(B_{ac}H_{bd} - 
H_{ac}B_{bd})\Omega^{cd}$ is the commutator between $B_{ab}$ and 
$H_{ab}$. This is the `real' version of the familiar 
relation $d\langle B\rangle/dt=i\langle[B,H]\rangle$. It is important 
to note that use of the `modified' Schr\"odinger equation does not 
affect this result. \par 

If $P_{ab}$ and $Q_{ab}$ are symmetric Hermitian operators satisfying 
the commutation relation  
\begin{equation} 
\left( P_{ac}Q_{bd} - Q_{ac}P_{bd}\right)\Omega^{cd}\ =\ g_{ab}\ , 
\label{eq:canc} 
\end{equation} 
then we say that $P_{ab}$ is {\it canonically conjugate} to $Q_{ab}$, 
and we refer to (\ref{eq:canc}) as the Heisenberg canonical 
commutation relation. This would apply, for example, when $P_{ab}$ 
and $Q_{ab}$ are the self-adjoint position and momentum operators of 
a quantum system. In fact, the Heisenberg commutation relation 
(\ref{eq:canc}) has to be regarded to some degree as formal, since 
the domain in ${\cal H}$ over which (\ref{eq:canc}) is valid is not 
necessarily obvious. This point can be remedied by consideration of 
the Weyl relation, which offers a more general and, ultimately, more 
rigourous basis for formulating the concept of canonical conjugacy. 
In real terms the Weyl relation is given by 
\begin{equation} 
\exp[-qJ^{a}_{\ b}P^{b}_{\ c}]\exp[pJ^{c}_{\ d}Q^{d}_{\ e}]\ =\ 
\exp[pJ^{a}_{\ b}(Q^{b}_{\ c}+q\delta^{b}_{\ c})]
\exp[-qJ^{c}_{\ d}P^{d}_{\ e}]\ , \label{eq:Weyl} 
\end{equation} 
where $p$ and $q$ are parameters. Note that in the Weyl relation 
the effect of interchanging the two terms on the left is to `shift' 
the operator $Q^{a}_{\ b}$ by the amount $q\delta^{a}_{\ b}$. The 
Heisenberg commutation relation (\ref{eq:canc}) is then obtained by 
formally differentiating (\ref{eq:Weyl}) with respect to $p$ and $q$, 
then setting them to zero. \par 

\section{Quantum Measurement} 

We shall now turn to the problem of parametric estimation for 
quantum mechanical states. By expressing quantum theory within 
a real Hilbert space framework, and studying the corresponding 
`real dynamics', we take advantage of the geometrical formulation 
of statistical inference outlined earlier. We begin with some 
remarks indicating the general setting for 
our investigations in quantum measurement and quantum statistical 
estimation. In particular, with a view to the parameter estimation 
problem we shall be considering shortly, it will suffice for our 
purposes to examine the case where we are concerned with the 
measurement of an observable with a continuous spectrum, such as 
position or momentum. \par 

We shall consider the situation where the system is in a pure 
state, characterised in real terms by a state vector $\xi^{a}$ in 
a real Hilbert space ${\cal H}$ equipped with an inner product 
$g_{ab}$ and a compatible complex structure $J^{a}_{\ b}$. It is 
possible also to consider the case where the state is described 
by a general density matrix, but this is not required for present 
purposes. \par 

The measurement of an observable is fully characterised in quantum 
mechanics by the specification of a {\it resolution of the 
identity}. By this we mean a one-parameter family $M_{ab}(x)$ of 
positive symmetric Hermitian operators that integrate up to form 
the identity operator. Thus we have $M_{ab}=M_{ba}$, 
$M_{ab}\xi^{a}\xi^{b}\geq0$ for any vector $\xi^{a}\in{\cal H}$, 
$M_{cd}J^{c}_{\ a}J^{d}_{\ b}=M_{ab}$, and 
\begin{equation} 
\int_{-\infty}^{\infty}M_{ab}(x)dx\ =\ g_{ab}\ . 
\end{equation} 
Then the probability that the observable $X$ represented by the 
measurement $M_{ab}(x)$ lies in the interval $\alpha <x<\beta$, if 
the state of the system is $\xi^{a}$, is given by 
\begin{equation} 
{\rm Prob}[\alpha<x<\beta]\ =\ 
\int_{\alpha}^{\beta}M_{ab}(x)\xi^{a}\xi^{b}dx\ , 
\label{eq:problaw} 
\end{equation} 
and for the expectation of $X$ we have 
\begin{equation} 
E_{\xi}[X]\ =\ \int_{-\infty}^{\infty} xM_{ab}(x) 
\xi^{a}\xi^{b}dx\ . 
\end{equation} 
The observable $X_{ab}$ itself, on the other hand, is given by 
\begin{equation} 
X_{ab}\ =\ \int_{-\infty}^{\infty} xM_{ab}(x)dx\ , 
\label{eq:spec} 
\end{equation} 
from which it follows that $E_{\xi}[X]=X_{ab}\xi^{a}\xi^{b}$. 
The probability law (\ref{eq:problaw}) is 
not readily ascertainable from the operator $X_{ab}$ directly, and 
this is why one needs the resolution of the identity $M_{ab}(x)$, 
or equivalently, the density function 
\begin{equation} 
p(x|\xi)\ =\ M_{ab}(x)\xi^{a}\xi^{b} 
\end{equation} 
for the random variable $X$, conditional on the specification of the 
state $\xi^{a}$. \par 

It is interesting to note the relationship between properties of the 
operator $X_{ab}$ defined by the spectral resolution (\ref{eq:spec}) 
and the corresponding resolution of the identity $M_{ab}(x)$. If 
$X_{ab}$ is a bounded operator, which is to say that there 
exists a constant $c$ such that $|X_{ab}\xi^{a}\xi^{b}|\leq c 
g_{ab}\xi^{a}\xi^{b}$ for all $\xi^{a}\in{\cal H}$, then there exists 
a unique spectral resolution (\ref{eq:spec}) with the following two 
additional properties: i) the resolution is {\it orthogonal}, or 
projection valued, in the sense that 
$g^{ab}M_{ac}(x)M_{bd}(y) = M_{cd}(x)\delta(x-y)$; and ii) the family 
of operators $M_{ab}(x)$ has compact support in the variable $x$. \par 

More generally, if $X_{ab}$ is self-adjoint (but not necessarily 
bounded) then there exists a unique orthogonal resolution of 
unity $M_{ab}(x)$ such that $X_{ab}$ is given by (\ref{eq:spec}), and 
that the domain of the operator $X_{ab}$ consists of all vectors 
$\xi^{a}$ satisfying 
\begin{equation} 
\int_{-\infty}^{\infty}x^{2}M_{ab}(x)\xi^{a}\xi^{b}dx
\ <\ \infty\ . \label{eq:s-a} 
\end{equation} 
On the other hand, if $X_{ab}$ is maximally symmetric, then there 
exists a unique resolution $M_{ab}(x)$ such that $X_{ab}$ has the 
spectral representation (\ref{eq:spec}), and the expectation of 
its square is given by 
\begin{equation} 
X^{c}_{\ a}X_{bc}\xi^{a}\xi^{b}\ =\ \int_{-\infty}^{\infty} 
x^{2}M_{ab}(x)\xi^{a}\xi^{b}dx 
\end{equation} 
for any state $\xi^{a}$ in the domain of $X_{ab}$, which is given 
by (\ref{eq:s-a}). In this case {\it the resolution of the identity 
is not orthogonal}. \par 

In noting these results we recall that the domain ${\cal D}(X)$ of 
a densely defined operator $X_{ab}$ consists of those state vectors 
$\xi^{a}$ for which $X_{ab}\xi^{b}$ exists, or equivalently, for which 
$X^{c}_{a}X_{bc}\xi^{a}\xi^{b}<\infty$. If $X_{ab}\xi^{a}\eta^{b} = 
X_{ab}\eta^{a}\xi^{b}$ for all $\xi^{a}, \eta^{a} \in {\cal D}(X)$, 
then we say $X_{ab}$ is symmetric, and write $X_{ab}=X_{ba}$. 
Then we define the adjoint domain ${\cal D}^{*}(X)$ to consist 
of all those vectors $\eta^{a}$ for which there exists a vector 
$\zeta^{a}$ such that $\eta^{a}(X_{ab}\xi^{b})=\zeta_{a}\xi^{b}$ for 
every $\xi^{a}$ in ${\cal D}(X)$. For any element $\eta^{a}\in 
{\cal D}^{*}(X)$ we thus define the adjoint operator $X^{*}_{ab}$, 
with domain ${\cal D}^{*}(X)$, by the action $X^{*}_{ab}\eta^{b} = 
\zeta_{a}$. If ${\cal D}(X)={\cal H}$, then 
$X_{ab}$ bounded; if ${\cal D}(X)={\cal D}^{*}(X)$ then we say 
$X_{ab}$ is self-adjoint. For a symmetric operator, 
${\cal D}(X) \subseteq {\cal D}^{*}(X)$. 
If ${\cal D}(X) \subseteq {\cal D}(Y)$ and if $X_{ab}=Y_{ab}$ on 
${\cal D}(X)$, then we say $Y_{ab}$ is an extension of 
$X_{ab}$. An operator is said to be maximally symmetric if 
it is symmetric, but has no self-adjoint extension. See, e.g., 
Bogolubov, et. al. (1975) or Reed and Simon (1974) for further 
details. \par 

Perhaps it can be stated that one of the most important modern 
developments in the understanding of basic quantum theory was the 
realisation that {\it general} measurements are given by positive 
operator-valued measures (POM), which involve general, nonorthogonal 
resolutions of the identity in an essential way. At the same time, 
one has to understand the category of observables in quantum 
mechanics to be widened on that basis to include maximally 
symmetric operators, as opposed to merely self-adjoint operators. 
See, e.g., Davies and Lewis (1970), Helstrom (1969), and 
Holevo (1973, 1979). \par 

We note this because some of the most interesting parameter 
estimation problems in quantum statistical inference involve 
nonorthogonal resolutions---for example, time and phase 
measurement---for which the relevant estimators are maximally 
symmetric operators characterised by nonorthogonal resolutions. \par 

In what follows we shall be particularly concerned with 
measurements associated with one-parameter families of unitary 
transformations. In this connection we point out that a 
transformation $\xi^{a}\rightarrow O^{a}_{\ b}\xi^{b}$ represents 
a general {\it rotation} of the real Hilbert space ${\cal H}$ 
about its origin if $O^{a}_{\ c}O^{b}_{\ d}g_{ab}=g_{cd}$. A 
unitary transformation $\xi^{a}\rightarrow U^{a}_{\ b}\xi^{b}$ is 
characterised by an operator $U^{a}_{\ b}$ that is both orthogonal, 
in the sense that the metric is preserved, and {\it symplectic}, 
in the sense that the symplectic structure $\Omega_{ab}$ is also 
preserved, so $U^{a}_{\ c}U^{b}_{\ d}\Omega_{ab}=\Omega_{cd}$. This 
gives us a characterisation of unitary transformations in purely 
real terms. \par 

Now suppose $U^{a}_{\ b}(\theta)$ is a continuous one-parameter 
family of unitary transformations on ${\cal H}$, satisfying 
$U^{a}_{\ c}(\theta)U^{c}_{\ b}(\theta')=U^{a}_{\ b}(\theta+\theta')$. 
Then there exists a self-adjoint 
operator $F_{ab}$ such that $U^{a}_{\ b}(\theta)=\exp[\theta 
J^{a}_{\ c}F^{c}_{\ b}]$. For example, if $P_{ab}$ is the momentum 
operator in a specific direction, then $U^{a}_{\ b}(\theta) = 
\exp[\theta J^{a}_{\ c}P^{c}_{\ b}]$ can be interpreted as a `shift' 
operator, and the one-parameter family of states $\xi^{a}(\theta) = 
U^{a}_{\ b}(\theta)\xi^{b}$ is obtained by shifting the states 
along the given axis. The question arises then, given such a family 
$\xi^{a}(\theta)$, what measurements can we perform to determine 
the relation of the true state of the system to the original state 
$\xi^{a}$? \par 

A resolution of the identity $M_{ab}(x)$ is said to be 
{\it covariant} with respect to the one-parameter family of unitary 
transformations $U^{a}_{b}(\theta)$ if 
\begin{equation} 
U^{c}_{\ a}(\theta)U^{d}_{\ b}(\theta)M_{cd}(x)\ =\ 
M_{ab}(x-\theta)\ . \label{eq:cov2} 
\end{equation} 
In that case it is straightforward to verify that the symmetric 
operator $\Theta_{ab}$ defined by 
\begin{equation} 
\Theta_{ab}\ =\ \int_{-\infty}^{\infty}(x-\mu)M_{ab}(x)dx\ , 
\end{equation} 
where $\mu=\Theta_{ab}\xi^{a}\xi^{b}/\xi_{c}\xi^{c}$, is an 
{\it unbiased estimator} for the parameter $\theta$ in the sense 
that 
\begin{equation} 
\frac{\Theta_{ab}\xi^{a}(\theta)\xi^{b}(\theta)}
{g_{cd}\xi^{c}(\theta)\xi^{d}(\theta)}\ =\ \theta 
\end{equation} 
for any state vector $\xi^{a}(\theta)$ along the specified 
trajectory. \par 

In particular, one can verify that if the symmetric 
operator $\Theta_{ab}$ is {\it canonically conjugate} to $F_{ab}$, 
then its spectral resolution $M_{ab}(x)$ necessarily satisfies the 
covariance relation (\ref{eq:cov2}). A symmetric operator 
$\Theta_{ab}$ is defined to be canonically conjugate to a 
self-adjoint operator $F_{ab}$ if for all values of the parameters 
$\theta,\phi$ we have 
\begin{equation} 
\exp[-\theta J^{a}_{\ b}F^{b}_{\ c}]\exp[\phi J^{c}_{\ d}
\Theta^{d}_{\ e}]\exp[\theta J^{e}_{\ f}F^{f}_{\ g}]\ =\ 
\exp[\phi J^{a}_{\ h}(\Theta^{h}_{\ g}+\theta\delta^{h}_{\ g})]\ . 
\label{eq:sccom}
\end{equation} 
In other words, the unitary transformation $U^{a}_{\ b}(\theta)$ 
has the effect of shifting $\Theta^{a}_{\ b}$ by the amount 
$\theta\delta^{a}_{\ b}$ in (\ref{eq:sccom}). If $F_{ab}$ and 
$\Theta_{ab}$ are self-adjoint, then the exponentials in 
(\ref{eq:sccom}) can be given meaning by a spectral representation, 
and (\ref{eq:sccom}) is equivalent 
to the Weyl relation. On the other hand, it may be that the given 
self-adjoint operator $F_{ab}$ has no self-adjoint canonical 
conjugate. Nevertheless there may exist a maximally symmetric 
operator $\Theta_{ab}$ satisfying (\ref{eq:sccom}), if we define  
\begin{equation} 
\exp[\phi J^{a}_{\ b}\Theta^{b}_{\ c}]\ :=\ \int_{-\infty}^{\infty} 
\left[ \cos(\phi x)\delta^{a}_{\ b} + \sin(\phi x)J^{a}_{\ b} 
\right] M^{b}_{\ c}(x)dx\ , 
\end{equation} 
where $M_{ab}(x)$ is the unique spectral resolution for $\Theta_{ab}$ 
satisfying the required conditions on its first and second moments. 
This occurs, for example, in the case of a Hermitian operator that is 
bounded below, which admits no self-adjoint canonical conjugate, but 
nevertheless under fairly general conditions admits a maximally 
symmetric canonical conjugate satisfying (\ref{eq:sccom}). \par 

Consider, for example, the case of a free particle in one dimension, 
for which the momentum and position operators are denoted $P$ and 
$Q$, and the Hamiltonian is $H=P^{2}/2m$. Then, $H$ has no 
self-adjoint canonical conjugate, but it does have a well defined 
maximally symmetric canonical conjugate, given (Holevo 1982) by 
\begin{equation} 
T\ =\ im\, {\rm sign}(P)|P|^{-1/2}Q|P|^{-1/2}\ , 
\end{equation} 
and it is not difficult to check, at least formally, that 
$TH-HT=i$. \par 

If we work in the usual momentum representation, for which the 
wave function is given by $\xi(p)$, assumed normalised, then 
${\cal D}(T)$ is given by those functions for which 
\begin{equation} 
\int_{-\infty}^{\infty}\left| \frac{d(\xi(p)|p|^{-1/2})}
{dp}\right|^{2} |p|^{-1} dp\ <\ \infty\ . 
\end{equation} 
If $\xi(p,t)$ is a one-parameter family of states satisfying the 
Schr\"odinger equation $i\partial_{t}\xi=H\xi$, we find that 
$\langle\xi(t)|T|\xi(t)\rangle=t+\langle\xi(0)|T|\xi(0)\rangle$, 
which shows that $T$ is an estimator for $t$. We mention this 
example to illustrate the point that even in a simple situation, the 
construction of the relevant estimator can be a subtle matter. \par 

\section{Quantum Estimation} 

Suppose now we consider a family of normalised state vectors 
$\xi^{a}(t)$, parameterised by the time $t$, that satisfy 
the Schr\"odinger equation (\ref{eq:sch2}). The curve $\xi^{a}(t)$ 
lies on the unit sphere ${\cal S}$ in the real Hilbert space 
${\cal H}$, and is characterised by the fact that it is the unique 
lift of the quantum mechanical state trajectory in the complex 
projective Hilbert space to the sphere ${\cal S}$ with the property 
that it is everywhere orthogonal to the direction $\zeta^{a} = 
J^{a}_{\ b}\xi^{b}$, as indicated in Proposition \ref{lem:22} 
(cf. figure 1). Regarding this curve as a statistical manifold, we 
shall study the problem of estimating the time parameter $t$ by use 
of the geometric techniques developed in \S\S 2-5. Let $T_{ab}$ 
denote an unbiased estimator for $t$. Thus $T_{ab}$ is a real 
symmetric Hermitian operator satisfying 
\begin{equation} 
\frac{T_{ab}\xi^{a}(t)\xi^{b}(t)}
{g_{cd}\xi^{c}(t)\xi^{d}(t)}\ =\ t\ , 
\label{eq:test}
\end{equation} 
for a system that is in the state $\xi(t)$. For example, if $T_{ab}$ 
is maximally symmetric and canonically conjugate to $H_{ab}$, then 
by the argument of the previous section we can make an adjustment of 
the form $T_{ab}\rightarrow T_{ab} + kg_{ab}$ for a suitable constant 
$k$ to remove the bias of $T_{ab}$, which does not change the 
conjugacy condition, and we are left with an estimator satisfying 
(\ref{eq:test}). \par 

Our idea is to apply the generalised Bhattacharyya bounds established 
in \S 5 to the quantum mechanical estimation problem, and consider 
the possibility of establishing sharper variants of the Heisenberg 
uncertainty relations $\Delta H\Delta T \geq 1/2$ in the case of 
canonically conjugate variables. \par 

The geometrical content of the generalised Bhattacharyya bound is 
that given the normal vector $\nabla_{a}t$ to the time-slice 
surfaces, we can choose a set of orthogonal vectors in ${\cal H}$ 
and express the length of this vector in terms of its orthogonal 
components. Then by use of Proposition \ref{prop:21} we can 
formulate a set of bounds on the variance of $T_{ab}$. In the 
`classical' setting for parameter estimation in 
\S 5, we found it natural to consider the orthogonal vectors given 
by $\hat{\xi}^{(k)}_{a}$ $(k=1, 2, \cdots)$, the $k$-th derivatives 
of the states $\xi_{a}$ projected orthogonally to the lower order 
derivatives. These satisfy ${\hat \xi}^{(j)}_{a}{\hat \xi}^{(k)}_{b}
g^{ab}=0$ for $j\neq k$, and will be referred to as the `classical 
system' of orthogonal vectors. In the quantum mechanical situation, the 
resulting scheme of possible sets of orthogonal vectors is somewhat 
richer, since the complex structure tensor can also be brought into
play. In particular, we find that the Cauchy-Riemann field 
$\zeta^{a}=J^{a}_{\ b}\xi^{b}$ is orthogonal to $\xi^{a}$, 
$\dot{\zeta}^{a}$ is orthogonal to $\dot{\xi}^{a}$, and so on. 
Therefore, we can construct a set of orthogonal vectors given in 
terms of $\xi^{a}, \zeta^{a}$, and their higher order derivatives. 
These will be referred to as the `quantum system' of orthogonal 
vectors, and denoted 
${\check \xi}^{(k)}_{a}$ and ${\check \zeta}^{(k)}_{a}$ 
$(k=0,1,2,\cdots)$, characterised by the property that 
${\check \xi}^{(j)}_{a}{\check \xi}^{(k)}_{a}g^{ab}=0$, 
${\check \xi}^{(j)}_{a}{\check \zeta}^{(k)}_{b}g^{ab}=0$, and 
${\check \zeta}^{(j)}_{a}{\check \zeta}^{(k)}_{b}g^{ab}=0$, for 
$j\neq k$. \par 

Before considering the higher order terms, we study the two lowest 
order terms arising from the quantum system, to note the familiar 
inequalities from standard quantum mechanics thus arising. In this 
case, the variance bound is: 
\begin{equation} 
{\rm Var}_{\xi}[T]\ \geq\ 
\frac{(\dot{\xi}^{a}\nabla_{a}t)^{2}}
{4\dot{\xi}^{b}\dot{\xi}_{b}} + 
\frac{(\dot{\zeta}^{a}\nabla_{a}t)^{2}}
{4\dot{\zeta}^{b}\dot{\zeta}_{b}} \ . \label{eq:heilb} 
\end{equation} 
\par 

\begin{guess} 
Let $\xi^{a}(t)$ satisfy the modified Schr\"odinger equation 
(\ref{eq:sch2}), and set $\zeta^{a}=J^{a}_{\ b}\xi^{b}$. Let $T_{ab}$ 
be an unbiased estimator for $t$. Then, 
$\dot{\xi}^{a}\nabla_{a}t=1$, and 
$\dot{\zeta}^{a}\nabla_{a}t = -2{\rm Cov}_{\xi}[H,T]$, where 
${\rm Cov}_{\xi}[H,T]$ denotes the covariance of the operators 
$H_{ab}$ and $T_{ab}$ in the 
state $\xi^{a}$. \label{prop:23} 
\end{guess} 

The proof is as follows. The fact that ${\dot \xi}^{a}\nabla_{a}t=1$ 
follows directly from the chain rule. Alternatively, notice that 
differentiation of (\ref{eq:test}) with respect to $t$ implies, by 
use of (\ref{eq:sch2}), that 
\begin{equation} 
\frac{2T_{ac}H_{bd}\Omega^{cd}\xi^{a}\xi^{b}}
{g_{ab}\xi^{a}\xi^{b}}\ =\ 1 \label{eq:test2} 
\end{equation} 
for any state vector on the specified trajectory $\xi^{a}(t)$. On the 
other hand, ${\dot \xi}^{a}\nabla_{a}t = J^{a}_{\ b}{\tilde H}^{b}_{\ c}
\xi^{c}\nabla_{a}t$ by (\ref{eq:sch2}), and 
\begin{equation} 
\nabla_{a}t\ =\ \frac{2{\tilde T}_{ab}\xi^{b}}
{g_{cd}\xi^{c}\xi^{d}}\ , \label{eq:nablat} 
\end{equation} 
which taken together with (\ref{eq:test2}) imply 
${\dot \xi}^{a}\nabla_{a}t=1$, as required. It follows then from 
the definition of $\zeta^{a}$ together with (\ref{eq:sch2}) and 
(\ref{eq:comp}) that ${\dot \zeta}=-{\tilde H}^{a}_{\ b}\xi^{b}$. 
Thus we have ${\dot \zeta}^{a}\nabla_{a}t = -2{\tilde H}_{ac}\xi^{a}
\xi^{b}/\xi_{c}\xi^{c}$, which by virtue of the definition of 
covariance given in \S 2 leads to the result ${\dot \zeta}^{a}
\nabla_{a}t=-2{\rm Cov}_{\xi}[H,T]$.  \par 

Therefore, if we write 
$\Delta_{\xi}T^{2}:=\langle\xi|(T^{2}-\langle T\rangle^{2})
|\xi\rangle$ for the variance of the estimator $T$, then, for 
the lowest order terms in the quantum variance bound, the inequality 
(\ref{eq:heilb}) reads 
\begin{equation} 
\Delta_{\xi}T^{2}\Delta_{\xi}H^{2}\ \geq\ \frac{1}{4} 
+ {\rm Cov}_{\xi}^{2}[H,T]\ . 
\label{eq:heilb2} 
\end{equation} 
In obtaining this result we have used the fact that the Fisher 
information in (\ref{eq:heilb}) is given by: 
${\cal G} = 4\dot{\xi}^{a}\dot{\xi}_{a} = 
4J^{a}_{\ c}J^{b}_{\ d}g_{ab}{\tilde H}^{c}_{\ e}
{\tilde H}^{d}_{\ f}\xi^{e}\xi^{f} = 4H^{c}_{\ a}H_{bc} 
\xi^{a}\xi^{b}-4(H_{ab}\xi^{a}\xi^{b})^{2}=4\Delta_{\xi}H^{2}$, 
where $\Delta_{\xi}H^{2}={\rm Var}_{\xi}[H]$ is the variance of 
the Hamiltonian (squared energy uncertainty) in the state $\xi^{a}$. 
In this way, we recover the standard `textbook' account of the 
uncertainty relations (see, e.g., Isham 1995). In particular, the 
second term on the right of (\ref{eq:heilb2}) is usually represented 
by an anticommutator, via the relation 
\begin{equation} 
{\rm Cov}_{\xi}[H,T]\ =\ \frac{1}{2}E_{\xi}[{\tilde H}{\tilde T} 
+ {\tilde T}{\tilde H}]\ , 
\end{equation} 
where ${\tilde H}=H-E_{\xi}[H]$ and ${\tilde T}=T-E_{\xi}[T]$. 
If we omit the second term in (\ref{eq:heilb2}) and keep the first 
term, corresponding to a quantum extension of the standard 
Cram\'er-Rao lower bound (\ref{eq:CRI0}), we find 
\begin{equation} 
\Delta_{\xi}T^{2}\Delta_{\xi}H^{2}\ \geq\ \frac{1}{4} \ . 
\label{eq:dTdH} 
\end{equation} 
The statistical interpretation of this result is as follows. 
Suppose we are told that at $t=0$ a quantum mechanical system is 
in the state $\xi^{a}(0)$, and it evolves subsequently according 
to the Schr\"odinger equation, with a prescribed Hamiltonian. Some 
time later we are presented with the system (or perhaps a large 
number of independent, identical copies of it), and we are 
required to make a measurement (or a set of identically designed 
measurements on all the copies) to determine $t$. The measurement 
is given by an observable $T_{ab}$ characterised by a nonorthogonal 
resolution of the identity $M_{ab}(x)$. The probability that the 
result $T$ of a given measurement lies in the range $(\alpha,\beta)$ 
is ${\rm Prob}[\alpha<T<\beta] = \int_{\alpha}^{\beta}
M_{ab}(x)\xi^{a}(t)\xi^{b}(t)dx$, 
and for the expectation of $T$ we have $E_{\xi(t)}[T]=t$. Thus, by 
averaging the results on all the copies we can approximate the 
value of $t$. The variance of $T$ is necessarily bounded from 
below, in accordance with (\ref{eq:dTdH}). On the other hand, 
the variance of the average of the results on $n$ copies is 
$n^{-1}\Delta_{\xi}T^{2}$. Hence, by making repeated measurements 
on different copies of the system we can improve the reliability 
of the estimate for $t$, despite the uncertainty principle. \par 

\section{Higher Order Quantum Variance Bounds} 

Some general remarks are in order concerning the relations 
(\ref{eq:heilb2}). We note that although the first term in 
(\ref{eq:heilb2}) is independent of the specific choice of 
estimator, the second term involving the covariance between 
$H_{ab}$ and $T_{ab}$ depends on the choice of $T_{ab}$. 
Hence this term is often dropped in the consideration of 
uncertainty relations, although in general the bound must be 
sharper than what we have in (\ref{eq:dTdH}). On the other hand, 
the reader may have observed that in deriving (\ref{eq:heilb2}) 
we have not, in fact, assumed that $T_{ab}$ is canonically 
conjugate to $H_{ab}$. We have merely assumed that $T_{ab}$ is 
an estimator for $t$, for the given trajectory $\xi^{a}(t)$, in 
accordance with (\ref{eq:test}). This is a weaker condition 
than canonical conjugacy, and thus it is legitimate to enquire 
whether, under the assumption of canonical conjugacy, it might 
be possible to derive bounds sharper than (\ref{eq:dTdH}), but 
nevertheless independent of the specific choice of estimator.  
Therefore, following the general approach outlined in \S 5, we 
propose to study contributions from higher order Bhattacharyya 
type corrections to the CR lower bound to search for such terms. 
What we find is that some of the corrections depend upon the 
choice of $T_{ab}$, while others do not. Those terms that are 
independent of the choice of the estimator contribute to a set 
of {\it generalised Heisenberg relations} for quantum 
statistical estimation. \par 

Before investigating details, we present some general results 
useful in obtaining higher order corrections. We assume that the 
state trajectory $\xi^{a}$ satisfies the dynamical equation 
(\ref{eq:sch2}). \par 

\begin{lem} 
Let $\xi^{(n)a}$ denote the $n$-th derivative of $\xi^{a}$ with 
respect to the time parameter $t$. Then, $g_{ab}\xi^{(n)a}\xi^{(n)b} 
= \langle {\tilde H}^{2n} \rangle$, where $\la {\tilde H}^{r} \ra$ 
denotes the $r$-th moment of the Hamiltonian about its mean. 
\label{lem:23} 
\end{lem} 
 
This follows directly by differentiation of the Schr\"odinger 
equation (\ref{eq:sch2}), and use of the Hermiticity condition for 
the metric $g_{ab}$. An example for $n=1$ is given by the expression 
${\cal G}=4g_{ab}\dot{\xi}^{a}\dot{\xi}^{b} = 
4\langle{\tilde H}^{2}\rangle$ for the Fisher information. \par 

\begin{lem} 
For a Schr\"odinger state $\xi^{a}$, the even 
moments of the Hamiltonian about its mean are independent of 
the time parameter. \label{lem:24} 
\end{lem} 

Indeed, since 
$\la {\tilde H}^{2n} \ra = g_{ab}\xi^{(n)a}\xi^{(n)b}$, 
we obtain $\partial_{t}\la {\tilde H}^{2n} \ra = 
2\xi^{(n)a}g_{ab}J^{a}_{\ c}{\tilde H}^{c}_{\ d}\xi^{(n)d}$, which 
vanishes. An elementary consequence of this result is that for 
arbitrary $n$ we have:  
\begin{equation} 
g_{ab}\xi^{(n+1)a}\xi^{(n)b}\ =\ 0. \label{eq:121} 
\end{equation} 
 
A remarkable result which is essential in finding higher order 
corrections that are independent of the choice of $T$ is the 
following. 

\begin{guess} 
Let $T_{ab}$ be canonically conjugate to $H_{ab}$, and hence an 
unbiased estimator for the parameter $t$. Then 
\begin{equation} 
T_{ab}\xi^{(n)a}\xi^{(n)b}\ =\ 
t g_{ab}\xi^{(n)a}\xi^{(n)b} + k \ , 
\end{equation} 
where $k$ is a constant. Thus for each $n$, ${\tilde T}_{ab}\xi^{(n)a}
\xi^{(n)b}$ is a constant of the motion along the Schr\"odinger 
trajectory. \label{prop:24} 
\end{guess} 

{\it Proof}. We recall from (\ref{eq:canc}) that $T_{ab}$ is 
canonically conjugate 
to $H_{ab}$ if $(T_{ac}H_{bd}-H_{ac}T_{bd})\Omega^{cd}=g_{ab}$, a 
relation which can also be written in the symmetric form 
\begin{equation} 
(T_{ac}H_{bd}+T_{bc}H_{ad})\Omega^{cd}\ =\ g_{ab}\ . 
\label{eq:canc3} 
\end{equation} 
Now suppose we differentiate $T_{ab}\xi^{(n)a}\xi^{(n)b}$. The 
Schr\"odinger equation in the form ${\dot \xi}^{a} = \Omega^{ac} 
{\tilde H}_{cd}\xi^{d}$ implies ${\dot \xi}^{(n)a} = \Omega^{ac} 
{\tilde H}_{cd}\xi^{(n)d}$, and thus 
\begin{equation} 
\partial_{t}(T_{ab}\xi^{(n)a}\xi^{(n)b})\ =\ 
2T_{ab}\Omega^{ac}{\tilde H}_{cd}\xi^{(n)d}\xi^{(n)b}\ . 
\label{eq:prop12} 
\end{equation} 
Since $T_{ab}\Omega^{ac}g_{cd}\xi^{(n)d}\xi^{(n)b}$ vanishes 
automatically, we can replace the ${\tilde H}_{cd}$ on the 
right-hand side of (\ref{eq:prop12}) with $H_{cd}$. However, 
according to (\ref{eq:canc3}) we have 
\begin{equation} 
2T_{ab}\Omega^{ac}{\tilde H}_{cd}\xi^{(n)d}\xi^{(n)b}\ =\ 
g_{bd}\xi^{(n)b}\xi^{(n)d}\ . 
\end{equation} 
On the other hand, Lemma \ref{lem:24} says that 
$g_{ab}\xi^{(n)a}\xi^{(n)b}$ is independent of $t$. Thus by 
integration of (\ref{eq:prop12}) we obtain the desired result.  \par 

\begin{lem} 
Let $T_{ab}$ be canonically conjugate to $H_{ab}$. Then for odd 
integers $n$, with $m=(n-1)/2$, we have:  
\begin{equation} 
2T_{ab}\xi^{a(n)}\xi^{b}\ =\ (-1)^{m} n 
g_{ab} \xi^{(m)a}\xi^{(m)b} \ . 
\end{equation} 
\label{lem:26} 
\end{lem} 

We sketch the derivation of this result. First, for $n=1$, 
it follows from $T_{ab}\xi^{a}\xi^{b}=t$ that 
$2T_{ab}\dot{\xi}^{a}\xi^{b} = 1$.  \par 

By differentiating this twice, we find 
$T_{ab}\stackrel{...\ }{\xi^{a}}\xi^{b}+3T_{ab}\ddot{\xi}^{a}
\dot{\xi}^{b} = 0$. On the other hand, it follows from 
Proposition \ref{prop:24} that 
$T_{ab}\dot{\xi}^{a}\dot{\xi}^{b} = t\dot{\xi}^{a}\dot{\xi}_{a} 
+ k$. Formula (\ref{eq:121}) then allows us to deduce that 
$2T_{ab}\ddot{\xi}^{a} \dot{\xi}^{b} = \dot{\xi}^{a}\dot{\xi}_{a}$. 
This gives us the desired result in the case $n=3$, namely:   
$2T_{ab}\stackrel{...\ }{\xi^{a}}\xi^{b} = 
-3\dot{\xi}^{a}\dot{\xi}_{a}$. \par 

If we differentiate 
$T_{ab}\xi^{a}\xi^{b}=t$ five times, we obtain 
$T_{ab}\xi^{(5)a}\xi^{b} = -5 T_{ab}\xi^{(4)a}
\dot{\xi}^{b} - 10T_{ab}\stackrel{...\ }{\xi^{a}}\ddot{\xi}^{b}$.  
Then differentiating $2T_{ab}\ddot{\xi}^{a} \dot{\xi}^{b} = 
\dot{\xi}^{a}\dot{\xi}_{a}$ twice, we find 
$T_{ab}\xi^{(4)a}\dot{\xi}^{b} = 
-3T_{ab}\stackrel{...\ }{\xi^{a}}\ddot{\xi}^{b}$, 
from which it follows that  
$T_{ab}\xi^{(5)a}\xi^{b} = 5 T_{ab}\stackrel{...\ }{\xi^{a}}
\ddot{\xi}^{b}$. 
However, since $T_{ab}\ddot{\xi}^{a}\ddot{\xi}^{b} 
= t \ddot{\xi}^{a}\ddot{\xi}_{a}+k$, we deduce by use of 
(\ref{eq:121}) that 
$2T_{ab}\stackrel{...\ }{\xi^{a}}\ddot{\xi}^{b} = 
\ddot{\xi}^{a}\ddot{\xi}_{a}$, 
and the desired result follows for $n=5$, namely:  
$2T_{ab}\xi^{(5)a}\xi^{b} = 5\ddot{\xi}^{a}\ddot{\xi}_{a}$. 
Higher order formulae can be deduced analogously. \par 

Armed with these results we are now in a position to deduce some 
higher order corrections to the Heisenberg relations for canonically 
conjugate observables. Again, we consider the measurement problem 
for the parameter $t$ in the case of a one-parameter family of state 
vectors $\xi^{a}(t)$ generated by the Schr\"odinger evolution 
(\ref{eq:sch2}), with a given Hamiltonian $H_{ab}$. The observable 
$T_{ab}$ is then taken to be a canonically conjugate unbiased 
estimator for the parameter $t$, in accordance with the theory 
developed in \S 10. \par 

First we consider the two second order corrections to the variance 
bound for $T_{ab}$ arising when $\nabla_{a}t$ is expanded in 
terms of the `quantum system' of orthogonal vectors, given by 
${\dot \xi}^{a}$, ${\dot \zeta}^{a}$, ${\check \xi}^{(2)a}$, and 
${\check \zeta}^{(2)a}$. The variance bound then takes the 
form 
\begin{equation} 
{\rm Var}_{\xi}[T]\ \geq\ 
\frac{({\dot \xi}^{a}\nabla_{a}t)}{4{\dot \xi}^{b}{\dot \xi}_{b}} + 
\frac{({\dot \zeta}^{a}\nabla_{a}t)}{4{\dot \zeta}^{b}{\dot \zeta}_{b}} 
+ \frac{(\check{\xi}^{(2)a}\nabla_{a}t)^{2}}
{4\check{\xi}^{(2)a}\check{\xi}^{(2)}_{a}} + 
\frac{(\check{\zeta}^{(2)a}\nabla_{a}t)^{2}}
{4\check{\zeta}^{(2)a}\check{\zeta}^{(2)}_{a}}\ , \label{eq:510} 
\end{equation}  
where $\check{\xi}^{(2)a}$ and $\check{\zeta}^{(2)a}$ are given, 
respectively, by 
\begin{equation} 
\check{\xi}^{(2)a}\ =\ \ddot{\xi}^{a} - 
\frac{(\ddot{\xi}^{b}\dot{\zeta}_{b})}
{\dot{\zeta}^{c}\dot{\zeta}_{c}} \dot{\zeta}^{a} - 
(\ddot{\xi}^{b}\xi_{b})\xi^{a} \label{eq:128} 
\end{equation} 
and 
\begin{equation} 
\check{\zeta}^{(2)a}\ =\ \ddot{\zeta}^{a} - 
\frac{(\ddot{\zeta}^{b}\dot{\xi}_{b})}
{\dot{\xi}^{c}\dot{\xi}_{c}} \dot{\xi}^{a} - 
(\ddot{\zeta}^{b}\zeta_{b})\zeta^{a} \ . 
\end{equation} 
Here we have used the fact that in order to obtain the second order 
(quantum) system of orthogonal vectors, we subtract the components 
of lower order derivatives of $\xi^{a}$ and $\zeta^{a}$ from 
$\ddot{\xi}^{a}$ and $\ddot{\zeta}^{a}$. There are only three terms 
appearing in these expressions since 
$\ddot{\xi}^{a}\dot{\xi}_{a}=\ddot{\xi}^{a}\zeta_{a}=0$ and 
$\ddot{\zeta}^{a}\dot{\zeta}_{a}=\ddot{\zeta}^{a}\xi_{a}=0$. It can 
be verified by use of (\ref{eq:comp}) and the Hermitian condition 
(\ref{eq:herm}) for $H_{ab}$ that the norms of $\check{\xi}^{(2)a}$ 
and $\check{\zeta}^{(2)a}$ agree, and are given by  
\begin{equation} 
\check{\xi}^{(2)a}\check{\xi}^{(2)}_{a}\ =\ 
\check{\zeta}^{(2)a}\check{\zeta}^{(2)}_{a}\ =\ 
\langle {\tilde H}^{2}\rangle^{2} K_{\xi}^{2}\ , 
\end{equation} 
where $K_{\xi}$ is defined by 
\begin{equation} 
K_{\xi}^{2}\ :=\ 
\frac{\la{\tilde H}^{4}\ra}{\la{\tilde H}^{2}\ra^{2}} 
- \frac{\la{\tilde H}^{3}\ra^{2}}
{\la{\tilde H}^{2}\ra^{3}} - 1\ . 
\end{equation} 
We note that $K_{\xi}^{2}$ is the curvature of the corresponding 
classical `thermal' state defined by the differential equation 
\begin{equation}
\frac{\partial\psi^{a}}{\partial\beta}\ =\ - \frac{1}{2} 
{\tilde H}^{a}_{\ b}\psi^{b}\ . \label{eq:thermo} 
\end{equation} 
The term `thermal state' used in this context is meant 
to suggest that we identify the parameter in the 
differential equation (\ref{eq:thermo}) with the inverse 
temperature $\beta$ (Brody and Hughston 1996d, 1997a). It is 
interesting to note that the Schr\"odinger trajectories 
form a family of curves orthogonal to the corresponding classical 
thermal state trajectories, i.e., wherever they meet we have 
$\dot{\psi}^{a}\dot{\xi}_{a}=0$. According to the argument outlined 
in \S 4, a thermal trajectory comprises an exponential family of 
distributions. However, since $\Omega_{ac}H^{c}_{\ b}$ is 
antisymmetric, the Schr\"odinger equation does 
not generate an exponential family in the $t$ variable. \par 

The first two terms on the right of (\ref{eq:510}) lead, as we 
have seen, to the standard first order uncertainty relation 
(\ref{eq:heilb2}). Now we proceed to value the second order 
terms. The numerators appearing in the second order terms in 
(\ref{eq:510}) can be calculated as follows. We consider the 
term involving $\hat{\xi}^{(2)a}$ first. By use of 
(\ref{eq:nablat}) and (\ref{eq:128}), and the fact that ${\check \xi}
^{(2)a}\xi^{b}g_{ab}=0$, this can be seen to be given by four 
times the square of the expression 
\begin{equation}    
\left( \ddot{\xi}^{a} - 
\frac{(\ddot{\xi}^{b}\dot{\zeta}_{b})}
{\dot{\zeta}^{c}\dot{\zeta}_{c}} \dot{\zeta}^{a} - 
(\ddot{\xi}^{b}\xi_{b})\xi^{a} \right) T_{ad}\xi^{d}\ . 
\label{eq:1213} 
\end{equation} 
Since $2T_{ab}\dot{\xi}^{a}\xi^{b}=1$, we have 
$T_{ab}\ddot{\xi}^{a}\xi^{b}=-T_{ab}\dot{\xi}^{a}\dot{\xi}^{b}$. 
Likewise ${\ddot \xi}^{a}\xi_{a}=-{\dot \xi}^{a}{\dot \xi}_{a}$. 
Thus for the first and third terms in (\ref{eq:1213}) we can write 
\begin{equation} 
T_{ab}{\ddot \xi}^{a}\xi^{b}-{\ddot \xi}^{c}\xi_{c}T_{ab}
\xi^{a}\xi^{b}\ =\ -{\tilde T}_{ab}{\dot \xi}^{a}{\dot \xi}^{b}\ , 
\end{equation} 
which, by Proposition \ref{prop:21}, is constant along the 
Schr\"odinger trajectory. As a consequence 
of the Hermitian condition on $T_{ab}$, we find that 
\begin{equation} 
{\tilde T}_{ab}\dot{\xi}^{a}\dot{\xi}^{b}\ =\ 
{\tilde H}^{a}_{\ c}{\tilde T}_{ab}
{\tilde H}^{b}_{\ d}\xi^{c}\xi^{d}\ =\ \{ 
{\tilde T},{\tilde H}^{2}\}_{\xi}\ , 
\label{eq:518} 
\end{equation} 
where $\{{\tilde T},{\tilde H}^{2}\}$ is the anticommutator of 
${\tilde T}$ and ${\tilde H}^{2}$. In deriving the second 
equality in (\ref{eq:518}) we make use of the identity 
$TH^{2}+H^{2}T = 2HTH$, which is valid for canonically conjugate 
operators. On the other hand, the dynamical equation for $\zeta^{a}$ 
implies that the expression $T_{ab}\dot{\zeta}^{a}\xi^{b}$ 
appearing in the second term of (\ref{eq:1213}) is minus the 
expectation of the anticommutator of ${\tilde T}$ and $\tilde{H}$. 
Since $\ddot{\xi}^{a}=-{\tilde H}^{a}_{\ b}{\tilde H}^{b}_{\ c}
\xi^{c}$ and $\dot{\zeta}^{a}=-{\tilde H}^{a}_{\ b}\xi^{b}$, we find 
that $\ddot{\xi}^{a}\dot{\zeta}_{a}=\langle{\tilde H}^{3}\rangle$ 
and $\dot{\zeta}^{a}\dot{\zeta}_{a}=\langle{\tilde H}^{2}\rangle$. 
Thus, combining together these various expressions, we obtain: 
\begin{equation} 
\frac{(\check{\xi}^{(2)a}\nabla_{a}t)^{2}}
{4\check{\xi}^{(2)a}\check{\xi}^{(2)}_{a}}\ =\ 
\frac{1}{4\la{\tilde H}^{2}\ra K^{2}_{\xi}}\left( 
\{ {\tilde T},{\tilde H}^{2}\}_{\xi} - 
\frac{\la{\tilde H}^{3}\ra}{\la{\tilde H}^{2}\ra} 
\{ {\tilde T},{\tilde H}\}_{\xi} \right)^{2} \ . \label{eq:519} 
\end{equation} 

Let us turn to the term in (\ref{eq:510}) involving 
$\check{\zeta}^{(2)a}$. We find that 
$T_{ab}\ddot{\zeta}^{a}\xi^{b} = 0$ and 
$T_{ab}\zeta^{a}\xi^{b} = 0$. Since 
$\ddot{\zeta}^{a}\dot{\xi}_{a}=-\langle{\tilde H}^{3}\rangle$, the 
contribution from the $\check{\zeta}^{(2)a}$ term is thus 
\begin{equation} 
\frac{(\check{\zeta}^{(2)a}\nabla_{a}t)^{2}}
{4\check{\zeta}^{(2)a}\check{\zeta}^{(2)}_{a}}\ =\ 
\frac{1}{4\la{\tilde H}^{2}\ra}\left( 
\frac{\la{\tilde H}^{3}\ra^{2}}
{\la{\tilde H}^{2}\ra^{3}K^{2}_{\xi}}\right) \ . \label{eq:520} 
\end{equation} 
If we omit the terms contributing from (\ref{eq:519}), 
which depend upon the features of the specific choice of 
estimator $T$, then by consideration of the terms represented 
in (\ref{eq:520}) we obtain the following sharpened variance 
bound for $T_{ab}$ which takes the form of a generalised 
Heisenberg relation: 

\begin{guess} 
If $T$ and $H$ are canonically conjugate, then the following 
bound applies to the product of their variances in the 
Schr\"odinger state $\xi^{a}(t)$: 
\begin{equation} 
\la{\tilde T}^{2}\ra\la{\tilde H}^{2}\ra\ \geq\ 
\frac{1}{4} \left( 1 + \frac{\la{\tilde H}^{3}\ra^{2}}
{\la{\tilde H}^{2}\ra^{3}K_{\xi}^{2}} \right)\ .  
\label{eq:secb} 
\end{equation} 
\label{prop:25} 
\end{guess} 

The inequality (\ref{eq:secb}) is expressed in terms of natural 
statistical `invariants', namely, the skewness 
$\la{\tilde H}^{3}\ra^{2}/\la{\tilde H}^{2}\ra^{3}$ 
and the curvature $K_{\xi}^{2}$. Alternatively, we can write 
(\ref{eq:secb}) directly in terms of the central moments of 
the Hamiltonian: 
\begin{equation} 
\la{\tilde T}^{2}\ra\la{\tilde H}^{2}\ra\ \geq\ 
\frac{1}{4} \left( 1 + \frac{\la{\tilde H}^{3}\ra^{2}} 
{\la{\tilde H}^{4}\ra\la{\tilde H}^{2}\ra - 
\la{\tilde H}^{3}\ra^{2} - \la{\tilde H}^{2}\ra^{3}} 
\right) \ . \label{eq:522} 
\end{equation} 
The positivity of the denominator in the correction term can be 
verified directly by noting that this is the squared norm of the 
state $|\psi\ra$ defined by 
\begin{equation} 
|\psi\ra\ =\ \left( {\tilde H}^{2} - 
\frac{\la{\tilde H}^{3}\ra}{\la{\tilde H}^{2}\ra}{\tilde H}  
- \la{\tilde H}^{2}\ra \right) |\xi\ra \ , 
\end{equation} 
assuming $\la\xi|\xi\ra=1$, which is nonvanishing providing that 
${\tilde H}^{2}|\xi\ra$ does not lie in the span of 
${\tilde H}|\xi\ra$ and $|\xi\ra$. This also follows from the 
statistical identity noted in connection with formula 
(\ref{eq:statcurv}). \par 

As a further illustration of the general formalism, we exhibit 
another, distinct bound on the variance, independent of the specific 
choice of estimator for the time parameter $t$, that arises 
naturally when we consider inequalities based on the `classical 
system' of orthogonal vectors associated with $\xi^{a}(t)$. This 
bound can be derived when we examine the third order Bhattacharrya 
type correction, which is given by $({\hat \xi}^{(3)a}
\nabla_{a}\tau)^{2}/4{\hat \xi}^{(3)a}{\hat \xi}^{(3)}_{a}$, where 
${\hat \xi}^{(3)a}$ is the component of $\xi^{(3)a}$ orthogonal to 
$\xi^{a}, \dot{\xi}^{a}$ and $\ddot{\xi}^{a}$. Now we know from Lemma 
\ref{lem:24} that $\dot{\xi}^{a}\dot{\xi}_{a}$ is constant along 
quantum trajectories, so $\ddot{\xi}^{a}\dot{\xi}_{a}=0$. Furthermore, 
$\ddot{\xi}^{a}\xi_{a}=-\dot{\xi}^{a}\dot{\xi}_{a}$, so 
$\xi^{(3)a}\xi_{a} = 0$. Likewise, since 
$\ddot{\xi}^{a}\ddot{\xi}_{a}$ is constant, we have 
$\xi^{(3)a}\ddot{\xi}_{a} = 0$. Thus 
$\xi^{(3)a}$ is automatically orthogonal to $\xi^{a}$ and 
$\ddot{\xi}^{a}$ along quantum trajectories. It follows that 
\begin{equation} 
{\hat \xi}^{(3)a}\ =\ \xi^{(3)a} - 
\left( \frac{\xi^{(3)}_{b}\dot{\xi}^{b}}
{\dot{\xi}_{c}\dot{\xi}^{c}} \right) \dot{\xi}^{a} \ . 
\end{equation} 
We are interested in the variance bound obtained by consideration 
of the first and third terms in (\ref{eq:gbb3}): 
\begin{equation} 
{\rm Var}_{\xi}[T]\ \geq\ \frac{(\dot{\xi}^{a}\nabla_{a}t)^{2}}
{4\dot{\xi}_{b}\dot{\xi}^{b}} + 
\frac{({\hat \xi}^{(3)a}\nabla_{a}t)^{2}}{4{\hat \xi}^{(3)}_{a}
{\hat \xi}^{(3)a}} \ . \label{eq:gbb31} 
\end{equation} 
Now, $\xi^{(3)a}\nabla_{a}t = -3\langle{\tilde H}^{2}\rangle$ 
according to Lemma \ref{lem:26}. On the other hand, we have 
$\xi^{(3)}_{a}\dot{\xi}^{a}=-\langle{\tilde H}^{4}\rangle$ and 
$\xi^{(3)}_{a}\xi^{(3)a}=\langle{\tilde H}^{6}\rangle$, from which 
it follows that 
\begin{equation} 
{\hat \xi}^{(3)}_{a}{\hat \xi}^{(3)a}\ =\ \langle{\tilde H}^{6}
\rangle - \frac{\langle{\tilde H}^{4}\rangle^{2}}
{\langle{\tilde H}^{2}\rangle} \ . 
\end{equation} 
Putting these ingredients together, we thus obtain the following 
correction to the Cram\'er-Rao lower bound (cf. Brody and 
Hughston 1996b,c). \par 

\begin{guess} 
If $T$ and $H$ are canonically conjugate observable variables, then 
the following inequality holds along the Schr\"odinger trajectory 
$\xi^{a}(t)$ generated by $H$: 
\begin{equation} 
\langle {\tilde T}^{2}\rangle 
\langle{\tilde H}^{2}\rangle\ \geq\ 
\frac{1}{4} \left( 1+ 
\frac{(\langle{\tilde H}^{4}\rangle - 3 \langle 
{\tilde H}^{2}\rangle^{2})^{2}}{\langle{\tilde H}^{6} 
\rangle\langle{\tilde H}^{2}\rangle - 
\langle{\tilde H}^{4}\rangle^{2}}\right)\ . \label{eq:fcum} 
\end{equation}
\label{prop:27} 
\end{guess} 

This correction is also strictly nonnegative, depends only 
on the given family of probability distributions determined by 
$\xi^{a}(t)$, and is independent of the specific choice of the 
estimator for time parameter. The fact that the denominator in 
the correction term is positive follows from the observation that 
it is given by $\langle{\tilde H}^{2}\rangle$ times the squared 
norm of the state $|\psi\rangle$ defined by 
\begin{equation} 
|\psi\rangle\ =\ \left( {\tilde H}^{3} - 
\frac{\langle{\tilde H}^{4}\rangle}{\langle{\tilde H}^{2}\rangle}
{\tilde H}\right) |\xi\rangle\ , 
\end{equation} 
where $\langle\xi|\xi\rangle=1$. It is interesting to note that 
the numerator in the correction 
is the square of the fourth cumulant of the distribution, 
usually denoted $\gamma_{2}$. The distributions for which 
$\gamma_{2}>0$ are called leptokurtic, and for $\gamma_{2}<0$ 
platykurtic. If the distribution is mesokurtic $(\gamma_{2}=0)$, 
then this correction vanishes, and an example of such a 
distribution is the Gaussian. For applications in quantum 
mechanics, we normally expect a distribution for $H$ that is 
not Gaussian, since $H$ is typically bounded from below, so 
(\ref{eq:fcum}) will generally give a nontrivial correction. 
In the case of other canonically conjugate variables, e.g., 
position and momentum, matters are different, and it is possible 
that a state can have a Gaussian distribution in these variables, 
as in the case of coherent states. \par 

In order to obtain some simple examples of the sort of numbers that 
might arise in connection with these corrections, suppose we 
assume that we have a physical system for which the energy is not 
definite, but rather has a known distribution, given by a density 
function $p(E)$. We shall examine the case when the energy has a 
gamma distribution, given by the density function of the form 
\begin{equation} 
p(E)\ =\ \frac{\sigma^{\gamma}}{\Gamma(\gamma)}
e^{-\sigma E}E^{\gamma-1}\ , \label{eq:expro} 
\end{equation} 
with $0\leq E\leq \infty$ and $\sigma, \gamma > 0$.  
This is to say, we have a large number of independent, identical 
systems with a prescribed Hamiltonian operator $H_{ab}$ and the 
Schr\"odinger state $\xi^{a}$. Then, by a set of measurements we can 
determine the distribution of the energy, which is characterised by 
the density function $p(E)$ given by 
\begin{equation} 
p(E)\ =\ \frac{1}{\sqrt{2\pi}}\int_{-\infty}^{\infty} 
\xi^{a}\xi_{b}\exp\left[ i\lambda(H^{b}_{\ a} - 
E\delta^{b}_{\ a})\right] d\lambda\ . 
\end{equation} 
For a given probability distribution for the energy, whether a 
self-adjoint Hamiltonian operator with the corresponding spectral 
resolution exists, or not, is an open problem which we hope to 
address elsewhere. Here, instead, we rely on the simple observation 
that the gamma distribution (\ref{eq:expro}) appears quite frequently 
in statistical studies, and hence it may help to provide an element 
of intuition as regards the behaviour of the correction terms. In the 
case of the gamma distribution, the moments are 
$ \la H^{n}\ra = (\gamma+n-1)!/\sigma^{n}(\gamma-1)!$, 
and for the corresponding lowest relevant 
central moments we find 
$\la{\tilde H}^{2}\ra=\gamma/\sigma^{2}$, 
$\la{\tilde H}^{4}\ra=3\gamma(\gamma+2)/\sigma^{4}$, and 
$\la{\tilde H}^{6}\ra=5\gamma(3\gamma^{2}+26\gamma+24)/\sigma^{6}$. 
It follows that the correction term in (\ref{eq:fcum}) 
is independent of the values of the parameter $\sigma$. 
We thus obtain  
\begin{equation} 
\langle {\tilde T}^{2}\rangle 
\langle{\tilde H}^{2}\rangle\ 
\geq\ \frac{1}{4} \left(1 + \frac{18}{3\gamma^{2} 
+ 47\gamma +42} \right)\ . 
\label{eq:secd} 
\end{equation}

In general, for Bhattacharyya style corrections based on the 
`classical system' of orthogonal vectors, the even order 
contributions turn out to be dependent upon the choice of the 
estimator $T$, while the odd order corrections are manifestly 
independent of the specific choice of $T$, and can be expressed 
entirely in terms of central moments of the conjugate observable $H$. 
For example, the fifth order correction can be shown to take the 
form (Brody and Hughston 1996c) 
\begin{equation}  
\frac{{\tilde H}^{2}[{\tilde H}^{8}
({\tilde H}^{4} - 3({\tilde H}^{2})^{2}) + 
{\tilde H}^{6}(8{\tilde H}^{4}{\tilde H}^{2} 
- {\tilde H}^{6}) - 5({\tilde H}^{4})^{3}]^{2}} 
{({\tilde H}^{10}({\tilde H}^{6}{\tilde H}^{2} 
- ({\tilde H}^{4})^{2}) + 2 {\tilde H}^{8} 
{\tilde H}^{6}{\tilde H}^{4} - 
({\tilde H}^{8})^{2}{\tilde H}^{2} - 
({\tilde H}^{6})^{3})({\tilde H}^{6}
{\tilde H}^{2} - ({\tilde H}^{4})^{2})} . \label{eq:thrd} 
\end{equation}  
Here we have used the slightly simplified notation ${\tilde H}^{n}$  
for the $n$-th moment of the Hamiltonian about its mean. If we assume 
that the distribution of the energy is given by a basic exponential 
distribution with probability density 
$p(E) = \sigma\exp(-\sigma E)$, which corresponds to the value 
$\gamma=1$ for the gamma distribution (\ref{eq:expro}), then 
the corrections (\ref{eq:fcum}) and (\ref{eq:thrd}) lead to the 
following bound, independent of the specific value of $\sigma$:  
\begin{equation} 
\langle {\tilde T}^{2}\rangle 
\langle{\tilde H}^{2}\rangle\ 
\geq\ \frac{1}{4} \left(1 + \frac{9}{46} + 
\frac{18,284,176}{290,027,815} \right)\ . \label{eq:25p} 
\end{equation}
\par 

The bounds given by (\ref{eq:fcum}) and (\ref{eq:thrd}) are 
significant inasmuch as they apply even if the odd-order central 
moments of the Hamiltonian vanish, in which case (\ref{eq:secb}) 
would no longer extend the standard Heisenberg relation. \par 

Throughout the discussion here we have confined the argument to 
consideration of the time measurement problem. In this case we 
consider the one-parameter family of states generated by the 
Hamiltonian. However, the same line of argument will apply for 
other pairs of canonically conjugate observables, such as 
position and momentum. \par 

The results indicated here can be pursued further in other ways 
as well, allowing us to consider various examples of natural 
statistical submanifolds of the quantum state space. For example, 
in a quantum field theoretic context it is natural to examine the 
coherent state submanifold of a bosonic Fock space. The geometry of 
this manifold arises when we consider measurements of the `classical' 
field associated with the POM generated by the family of all 
coherent states. Another interesting line of investigation intimately 
related to the arguments considered here concerns the status of 
thermodynamic states in classical and quantum statistical mechanics 
(Brody and Hughston 1996d, 1997a,b). \par 

\begin{acknowledgments} 
The authors gratefully acknowledge R. Brenner, T.R. Field, 
G.W. Gibbons, C.D. Hill, C.J. Isham, T.W.B. Kibble, 
B.K. Meister, A.F.S. Mitchell, R. Penrose, N. Rivier, 
R.F. Streater, and G.S. Watson for stimulating discussions. \par 
\end{acknowledgments} 

$*$ Electronic address: d.brody@damtp.cam.ac.uk \par 
$\dagger$ Electronic address: lane@ml.com\par 



\end{document}